\documentclass[11pt]{article}
\pdfoutput=1

\usepackage{epsfig}
\usepackage{psfrag}
\usepackage{latexsym}
\usepackage{indentfirst}
\usepackage{fancyhdr}
\usepackage{dsfont}
\usepackage{amsmath}
\usepackage{amssymb}
\usepackage{amsfonts}
\usepackage{mathrsfs}
\usepackage{amsthm}
\usepackage{pifont}
\usepackage{dsfont}
\usepackage{multirow}
\usepackage{array}
\usepackage{chngpage}
\usepackage{longtable}
\usepackage{cite}
\usepackage{bbold}
\usepackage{color}
\usepackage{braket}
\usepackage{colordvi}
\usepackage{fancybox}
\usepackage[footnotesize]{caption2}
\usepackage{graphicx}
\usepackage[center,footnotesize,hang]{subfigure}
\usepackage{bbm}
\usepackage[colorlinks, linkcolor=red, anchorcolor=black, citecolor=green]{hyperref}
\usepackage{multirow}
\usepackage{array}
\usepackage{textcomp}  %  for the command \textlbrackdbl
\usepackage{enumitem}
\usepackage{mathrsfs}  % $\mathscr{ABCDEFGHIJKLMNOPQRSTUVWXYZ}$
\newcommand{\PreserveBackslash}[1]{\let\temp=\\#1\let\\=\temp}
\newcolumntype{C}[1]{>{\PreserveBackslash\centering}p{#1}}
\newcolumntype{R}[1]{>{\PreserveBackslash\raggedleft}p{#1}}
\newcolumntype{L}[1]{>{\PreserveBackslash\raggedright}p{#1}}
\allowdisplaybreaks  % automated pagination

\newcommand{\bq}{\begin{eqnarray}}
\newcommand{\nq}{\end{eqnarray}}

\def\be{\begin{equation}}
\def\ee{\end{equation}}

\begin{document}

\title{\hfill ~\\[-30mm] \hfill\mbox{\small USTC-ICTS-19-02}\\[10mm]
        \textbf{{\Large Dihedral flavor group as the key to understand quark and lepton flavor mixing    }  }}

\date{}

\author{\\[1mm]Jun-Nan Lu\footnote{Email: {\tt hitman@mail.ustc.edu.cn}}~,~Gui-Jun Ding\footnote{Email: {\tt dinggj@ustc.edu.cn}}\\ \\
\it{\small Interdisciplinary Center for Theoretical Study and  Department of Modern Physics, }\\
\it{\small University of Science and
    Technology of China, Hefei, Anhui 230026, China}\\[4mm] }
\maketitle

\begin{abstract}
\noindent

We have studied the lepton and quark mixing patterns which can be derived from the dihedral group $D_n$ in combination with CP symmetry. The left-handed lepton and quark doublets are assigned to the direct sum of a singlet and a doublet of $D_n$. A unified description of the observed structure of the quark and lepton mixing can be achieved if the flavor group $D_n$ and CP are broken to $Z_2\times CP$ in neutrino, charged lepton, up quark and down quark sectors, and the minimal group is $D_{14}$. We also consider another scenario in which the residual symmetry of the charged lepton and up quark sector is $Z_2$ while $Z_2\times CP$ remains preserved by the neutrino and down quark mass matrices. Then $D_7$ can give the experimentally favored values of CKM and PMNS mixing matrices.

\end{abstract}
\thispagestyle{empty}
\vfill

\newpage
\setcounter{page}{1}

\section{Introduction}

It is well established that the three generations of quarks are mixed with each other to form mass eigenstates in the standard model. The quark flavor mixing matrix appearing in the weak charged-current interactions is referred to as the Cabibbo-Kobayashi-Maskawa (CKM) matrix~\cite{Tanabashi:2018oca}. The quark mixing angles exhibit a strongly hierarchical structure, and the largest one is the Cabibbo mixing angle $\theta_c\simeq13^{\circ}$ between the first and the second generation. Observation of neutrino oscillation implies that neutrinos have masses and non-zero mixing. Analogously there should be a lepton mixing matrix in the weak charged-current interactions, and it is usually called Pontecorvo-Maki-Nakagawa-Sakata (PMNS) matrix~\cite{Tanabashi:2018oca}. However, the lepton mixing angles are less hierarchical, both solar and atmospheric mixing angles are large and the reactor angle $\theta_{13}\simeq8.61^{\circ}$ is of a similar size to the Cabibbo angle~\cite{Esteban:2018azc}.
As regards the CP violation phase, the CP violation in the quark sector has been precisely measured in a variety of meson decay processes. It is confirmed that the single complex phase in the CKM matrix is the dominant source of CP violation, and the angle $\alpha$ in the unitary triangle is determined to be $\alpha=(84.5^{+5.9}_{-5.2})^{\circ}$~\cite{Tanabashi:2018oca}. If neutrinos are Majorana particles, there are additional sources of CP violation in the lepton sector, e.g., the Majorana phases in the PMNS matrix. At present CP conservation in neutrino oscillation is disfavored at $2\sigma$ level, and the exact values of the leptonic Dirac CP phase $\delta_{CP}$ is unknown although weak evidence for $\delta_{CP}$ around $3\pi/2$ is reported by T2K~\cite{Abe:2017vif} and NO$\nu$A~\cite{NOvA:2018gge}. Non-zero $\delta_{CP}$ is also preferred by global analysis of neutrino oscillation data~\cite{Esteban:2018azc,deSalas:2017kay,Capozzi:2018ubv}.

The origin of the above flavor mixing structures of quarks and leptons is one of the most important problems in standard model. Many proposals have been advanced to explain this puzzle in the literature. In particular, the non-abelian discrete flavor symmetries
appear to be particularly suitable to reproduce the large flavor mixing angles of the leptons.
In this paradigm, the three generations of left-handed lepton doublets are usually assumed to transform as a three-dimensional representation of the discrete flavor group $G_f$ which is subsequently broken to different subgroups in the charged lepton and neutrino sectors, the mismatch between the two subgroups allows one to predict the PMNS matrix up to permutations of rows and columns.
In a similar way, the mismatch of residual symmetries in up and down quark sectors can be employed to determine the CKM mixing matrix~\cite{Lam:2007qc,Blum:2007jz,Holthausen:2013vba,Araki:2013rkf,Yao:2015dwa,Varzielas:2016zuo}. However, no finite group has been found that can predict the correct values of the three different quark mixing angles at leading order, and only the Cabibbo angle can be generated~\cite{Yao:2015dwa,Varzielas:2016zuo}.

A recent progress is to extend the discrete flavor symmetry with CP symmetry~\cite{Feruglio:2012cw,Holthausen:2012dk,Chen:2014tpa}. This approach turns out to be quite powerful and it allows for precise predictions of both lepton mixing angles and CP violating phases. It can lead to very predictive scenarios, where all the mixing angles and CP phases are related to a small number of input parameters~\cite{Chen:2014wxa,Chen:2015nha,Everett:2015oka,Everett:2016jsk,Chen:2015siy,Chen:2016ica,Chen:2018lsv,Chen:2018eou,Chen:2018zbq}. Many models and analyses of CP and flavor symmetries have been studied so far, e.g. $A_4$~\cite{Ding:2013bpa,Li:2016nap}, $S_4$~\cite{Ding:2013hpa,Feruglio:2013hia,Li:2013jya,Li:2014eia,Lu:2016jit,Penedo:2017vtf}, $\Delta(27)$~\cite{Branco:2015hea,Branco:2015gna}, $\Delta(48)$~\cite{Ding:2013nsa,Ding:2014hva}, $A_5$~\cite{Li:2015jxa,DiIura:2015kfa,Ballett:2015wia,Turner:2015uta,DiIura:2018fnk}, $\Delta(96)$~\cite{Ding:2014ssa} and the infinite group series $\Delta(3n^2)$~\cite{Hagedorn:2014wha,Ding:2015rwa}, $\Delta(6n^2)$~\cite{Hagedorn:2014wha,Ding:2014ora} and $D^{(1)}_{9n, 3n}$~\cite{Li:2016ppt}. Flavor and CP symmetries can also constrain the CP violation in leptogenesis~\cite{Chen:2016ptr,Hagedorn:2016lva,Li:2017zmk}. Moreover, after including CP symmetry, we can achieve a unified description of quark and lepton flavor mixing from a single flavor symmetry group
if the residual symmetries of the charged lepton, neutrino, up quark and down quark sectors are different $Z_2\times CP$ subgroups, and the minimal flavor symmetry is $\Delta(294)$~\cite{Li:2017abz}.
One could reduce the residual subgroups of the charged lepton and up quark sectors to $Z_2$ while distinct $Z_2\times CP$ residual symmetries remain preserved by the neutrino and up quark mass matrices. Then either PMNS or CKM matrices depend on only three real free parameters, and the $\Delta(294)$ is still the smallest flavor symmetry group to generate the
the experimentally preferred quark and lepton mixing patterns~\cite{Lu:2018oxc}. There are other proposals to explain lepton and quark mixing from flavor symmetry and CP symmetry, see~\cite{Lu:2018oxc,Hagedorn:2018gpw,Hagedorn:2018bzo} for different perspectives. We observe that the group order of the required flavor symmetry is a bit larger.

The dihedral group $D_n$ with general $n$ is the group of symmetries of a regular polygon and it doesn't have irreducible three-dimensional representation. It is found that the phenomenologically acceptable Cabibbo angle can be accommodated by the dihedral group $D_7$~\cite{Lam:2007qc,Blum:2007jz}. In the present work, we shall extend the dihedral flavor group to involve also CP as symmetry\footnote{The interplay of $D_3$ and CP has been studied in~\cite{Ma:2017trv}.}. Both left-handed lepton doublets and quark doublets are assumed to transform as a reducible three-dimensional representation which is the direct sum of a singlet and a doublet representation of $D_n$. We shall analyze the mixing patterns for leptons and quarks arising from the breaking of $D_n$ and CP symmetry into $Z_2\times CP$ in all the relevant quark and lepton sectors. It is remarkable that the $D_{14}$ group of order $28$ can give an acceptable prediction for the PMNS and
the CKM matrices at leading order. The possible mixing patterns are also studied for the second scenario in which the residual symmetries of the charged lepton and up quark mass matrices are reduced to $Z_2$.
Then we find that both PMNS and CKM mixing matrices can be accommodated by the dihedral group $D_7$.

The structure of the paper is as follows: in section~\ref{sec:general_approach} we show the general constraints on the mass matrix and how to extract the mixing matrix if a residual symmetry $Z^{g_{\psi}}_2\times X_{\psi}$ or $Z^{g_{\psi}}_2$ is preserved by the fermion fields $\psi\in\left\{\nu, e, u ,d\right\}$. In section~\ref{sec:dihedral_group_theory} we present the mathematical properties of the dihedral group $D_n$ which is employed as flavor symmetry, and the CP transformations compatible with the $D_n$ flavor group are discussed. In section~\ref{sec:Z2CPZ2CP} we analyze the predictions for lepton and quark mixing if the flavor symmetry $D_n$ and CP symmetry are broken to different $Z_2\times CP$ subgroups in the charged lepton (up quark) and neutrino (down quark) sectors. In section~\ref{sec:Z2Z2CP} we study the scenario that the residual symmetries of charged lepton and up quark sectors are $Z_2$ subgroups while the neutrino and down quark mass matrices are invariant under $Z_2\times CP$. Finally we summarize our main results and conclude in section~\ref{sec:summary}.

\section{\label{sec:general_approach}Lepton and quark mixing from residual symmetry}

In the following, we briefly review how the lepton mixing can be predicted from a flavor symmetry group $G_f$ and a CP symmetry which are broken down to two different subgroups of the structure $Z_2\times CP$ in the charged lepton and neutrino sectors~\cite{Lu:2016jit,Li:2017abz,Lu:2018oxc}. The quark CKM mixing matrix can be derived in an analogous way using this method. We assume that the three generations of left-handed leptons doublets $L\equiv(\nu, e)^{T}$ and quark doublets $Q\equiv(u, d)^{T}$ transform as a three-dimensional representation $\rho$ of $G_{f}$. Notice that the following results for mixing matrix hold true no matter whether $\rho$ is a reducible or irreducible representation of $G_{f}$. For each type of fermionic field $\psi\in\left\{\nu, e, u ,d\right\}$, the corresponding residual symmetry is denoted as $Z^{g_{\psi}}_2\times X_{\psi}$, where $g_{\psi}$ is the generator of the residual flavor symmetry $Z^{g_{\psi}}_2$ and it fulfills $g^2_{\psi}=1$. The residual CP transformation $X_{\psi}$ should be a symmetric and unitary matrix otherwise the mass spectrum would be partially or completely degenerate~\cite{Chen:2014wxa,Chen:2015nha}. The restricted consistency condition between the residual flavor and CP symmetries reads~\cite{Feruglio:2012cw,Ding:2013hpa,Chen:2014wxa,Chen:2015nha,Everett:2015oka,Everett:2016jsk}
\begin{equation}
X_{\psi}\rho^{*}(\psi)X^{-1}_{\psi}=\rho(\psi)\,.
\end{equation}
 Under the action of residual symmetry, the field $\psi$ transforms as
\begin{equation}
\psi(x)\stackrel{G_f}{\longmapsto}
\rho(g_{\psi}) \psi(x),
\qquad \psi(x)\stackrel{CP}{\longmapsto} i X_\psi \gamma ^0 \mathcal{C} \bar{\psi}^T(\mathcal{P}x)\,,
\end{equation}
where $\mathcal{C}$ is the charge conjugation matrix with $\mathcal{P}x=(t, -\vec{x})$. For the residual symmetry to hold, the mass matrix should satisfy the following conditions
\begin{eqnarray}
\label{eq:constraint_Dirac}\rho^{\dagger}(g_{\psi})m^{\dagger}_{\psi}m_{\psi}\rho(g_{\psi})=m^{\dagger}_{\psi}m_{\psi},\qquad X^{\dagger}_{\psi}m^{\dagger}_{\psi}m_{\psi}X_{\psi}=(m^{\dagger}_{\psi}m_{\psi})^{*}\,,
\end{eqnarray}
if $\psi$ is Dirac field, where the mass matrix $m_{\psi}$ is defined in the right-left basis $\psi^{c}m_{\psi}\psi$. If $\psi$ (neutrino in the standard model) is Majorana field, the invariance conditions are
\begin{equation}
\label{eq:constraint_Maj}\rho^{T}(g_{\psi})m_{\psi}\rho(g_{\psi})=m_{\psi},\qquad  X^{T}_{\psi}m_{\psi}X_{\psi}=m^{*}_{\psi}\,,
\end{equation}
where the Majorana mass matrix $m_{\psi}$ is defined as $\frac{1}{2}\psi^{T}m_{\psi}\psi$. We can diagonalize the mass matrices $m^{\dagger}_{\psi}m_{\psi}$ and $m_{\psi}$ with a unitary transformation $U_{\psi}$,
\begin{equation}
U^{\dagger}_{\psi}m^{\dagger}_{\psi}m_{\psi}U_{\psi}=\mathrm{diag}(m^2_1, m^2_2, m^2_3),\qquad U^{T}_{\psi}m_{\psi}U_{\psi}=\mathrm{diag}(m_1, m_2, m_3)\,.
\end{equation}
From Eqs.~(\ref{eq:constraint_Dirac}, \ref{eq:constraint_Maj}) we can derive that the residual symmetry $Z^{g_{\psi}}_2\times X_{\psi}$ leads to the following constraints on the unitary matrix $U_{\psi}$~\cite{Lu:2016jit,Li:2017abz,Lu:2018oxc},
\begin{equation}
\label{eq:constr_resF}U^{\dagger}_{\psi}\rho(g_{\psi})U_{\psi}=\pm P^{T}_{\psi}\mathrm{diag}(1, -1, -1) P_{\psi}
\end{equation}
and
\begin{eqnarray}
\label{eq:constr_resCP}U^{\dagger}_{\psi}X_{\psi}U^{*}_{\psi}\equiv Q^2_{\psi}=\left\{\begin{array}{cc}
\mathrm{diag}(e^{i\alpha_1}, e^{i\alpha_2}, e^{i\alpha_3}), & \textnormal{for Dirac field}~\psi\,,
\\[0.1in]
\mathrm{diag}(\pm1, \pm1, \pm1), & \textnormal{ for Majorana field}~\psi\,,
\end{array}
\right.
\end{eqnarray}
where $\alpha_{1,2,3}$ are real free parameters, $P_{\psi}$ is a generic three dimensional permutation matrix and it can take six possible forms generated by
\begin{equation}\label{eq:permutation}
P_{12}=\begin{pmatrix}
0  &  1  &  0 \\
1  &  0  &  0 \\
0  &  0  &  1
\end{pmatrix}, \quad\quad
P_{13}=\begin{pmatrix}
0    &   0    &   1  \\
0    &   1    &   0  \\
1    &   0    &   0
\end{pmatrix}, \quad\quad
P_{23}=\begin{pmatrix}
1  &  0 &  0 \\
0  &  0 &  1 \\
0  &  1 &  0
\end{pmatrix}\,.
\end{equation}
As shown in~\cite{Lu:2016jit,Li:2017abz,Lu:2018oxc,Yao:2016zev}, one can solve the constraint equations of Eqs.~(\ref{eq:constr_resF},\ref{eq:constr_resCP}) by performing Takagi factorization for the residual CP transformation $X_{\psi}$ with the properties
\begin{equation}
\label{eq:Takagi}X_{\psi}=\Sigma_{\psi}\Sigma^{T}_{\psi},\qquad  \Sigma^{\dagger}_{\psi}\,\rho(g_{\psi})\Sigma_{\psi}=\pm\textnormal{diag}(1, -1, -1)\,.
\end{equation}
Then the residual symmetry $Z^{g_{\psi}}_2\times X_{\psi}$ would determine the unitary transformation $U_{\psi}$ to be of the form~\cite{Lu:2016jit,Li:2017abz,Lu:2018oxc,Yao:2016zev}
\begin{equation}
\label{eq:Upsi}U_{\psi}=\Sigma_{\psi}S_{32}(\theta_{\psi})P_{\psi}Q^{\dagger}_{\psi}\,,
\end{equation}
with
\begin{equation}
S_{23}(\theta_{\psi})=\left(
\begin{array}{ccc}
1 &~   0  &~  0 \\
0  &~ \cos\theta_{\psi}   &~ \sin\theta_{\psi}  \\
0 &~  -\sin\theta_{\psi}  &~ \cos\theta_{\psi}
\end{array}
\right)\,,
\end{equation}
where the free parameter $\theta_{\psi}$ can be limited in the range $0\leq\theta_{\psi}<\pi$ without loss of generality. Since both quark and charged lepton carry non-zero electric charges, only neutrinos in standard model can be Majorana particle. From Eq.~\eqref{eq:Upsi} we can see that the difference between Dirac and Majorana neutrinos is in the diagonal matrix $Q_{\psi}$. In the case of Dirac neutrinos, $Q_{\psi}$ is a general phase matrix and it can be rotated away by appropriate field redefination. For Majorana neutrinos, the diagonal entries of $Q_{\psi}$ are $\pm1$ and $\pm i$ and the Majorana phases can be predicted up to $\pi$.

If only a residual flavor symmetry $Z^{g_{\psi}}_2$ is preserved by the mass matrix of $\psi$, the unitary transformation $U_{\psi}$ would be only subject to the constraint in Eq.~\eqref{eq:constr_resF}. Since $g_{\psi}$ is of order two and its representation matrix $\rho(g_{\psi})$ has two degenerate eigenvalues, the residual flavor symmetry $Z^{g_{\psi}}_2$ can only distinguish  one generation from the other two ones. As a result, only one column of $U_{\psi}$ is fixed and it takes the following form~\cite{Lu:2018oxc}
\begin{equation}
\label{eq:Upsi_Z2}
U_{\psi}=\Sigma_{\psi}U_{23}^{\dagger}(\theta_{\psi},\delta_{\psi})P_{\psi}Q_{\psi}^{\dagger}\,,
\end{equation}
where $\Sigma_{\psi}$ diagonalizes $\rho(g_{\psi})$ as in Eq.~\eqref{eq:Takagi}, $Q_{\psi}$ is an arbitrary diagonal phase matrix, and $U_{23}(\theta_{\psi}, \delta_{\psi})$ is a block diagonal unitary rotation in the (23)-plane with
\begin{equation}
U_{23}(\theta_{\psi},\delta_{\psi})=\begin{pmatrix}
1 & 0 & 0 \\
0 & \cos\theta_{\psi} & \sin\theta_{\psi} \\
0 & -\sin\theta_{\psi} & \cos\theta_{\psi} \end{pmatrix}
\begin{pmatrix}
1 & 0 & 0 \\
0 & e^{i\delta_{\psi}} & 0 \\
0 & 0 & e^{-i\delta_{\psi}} \end{pmatrix}\,,
\end{equation}
where the angles $\theta_{\psi}$ and $\delta_{\psi}$ need only be defined over the intervals $[0, \pi/2]$ and $[0, \pi)$ respectively. In comparison with the residual symmetry $Z^{g_{\psi}}_2\times X_{\psi}$, the Majorana phases cannot be predicted for Majorana neutrinos in this case.

As a result, if the flavor and CP symmetries are broken to $Z^{g_{\nu}}_2\times X_{\nu}$ and $Z^{g_{e}}_2\times X_{e}$ in the neutrino and charged lepton sectors respectively, the lepton mixing matrix would be given by
\begin{equation}
\label{eq:PMNS_Z2CPZ2CP}
U\equiv U^{\dagger}_{e}U_{\nu}=Q_{e}P^{T}_{e}S^{T}_{23}(\theta_{e})\Sigma^{\dagger}_{e}\Sigma_{\nu} S_{23}(\theta_{\nu})P_{\nu}Q^{\dagger}_{\nu}\,.
\end{equation}
In this scenario, all the mixing angles and CP phases only depend on two real rotation angles $\theta_{e}$ and $\theta_{\nu}$ in the interval between $0$ and $\pi$. If neutrinos are Majorana particles, $Q_{\nu}$ is a diagonal matrix with elements $\pm1$ and $\pm i$, and without loss of generality it can be parameterized as
\begin{equation}
\label{eq:Qnu}Q_{\nu}=\text{diag}(1, i^{k_1}, i^{k_2})\,,
\end{equation}
with $k_{1,2}=0,1,2,3$.
Analogously the residual symmetries $Z^{g_{u}}_2\times X_{u}$ and $Z^{g_{d}}_2\times X_{d}$ in the up type quark and down type quark sectors allow us to pin down the CKM mixing matrix as
\begin{equation}
\label{eq:CKM_Z2CPZ2CP}
V\equiv U^{\dagger}_{u}U_{d}=Q_{u}P^{T}_{u}S^{T}_{23}(\theta_{u})\Sigma^{\dagger}_{u}\Sigma_{d} S_{23}(\theta_{d})P_{d}Q^{\dagger}_{d}\,.
\end{equation}
In the following, we shall consider a second scenario in which the residual symmetry of the charged lepton sector is degraded to $Z^{g_e}_2$ such that the neutrino and charged lepton mass matrices exhibit the residual symmetries $Z^{g_e}_2$ and $Z^{g_{\nu}}_2\times X_{\nu}$ respectively. Then the lepton mixing matrix is given by
\begin{equation}
\label{eq:PMNS_Z2CPZ2}
U=Q_{e}P^{T}_{e}U_{23}(\theta_{e},\delta_{e})\Sigma_{e}^{\dagger}
\Sigma_{\nu}S_{23}(\theta_{\nu})P_{\nu}Q^{\dagger}_{\nu}\,,
\end{equation}
which depends on three free parameters $\theta_{e}$, $\delta_{e}$ and $\theta_{\nu}$. In the same fashion, if the residual symmetries $Z^{g_u}_2$ and $Z^{g_{d}}_2\times X_{d}$ are preserved in the up and down quark sectors respectively, the quark mixing matrix would be determined by
\begin{equation}
\label{eq:CKM_Z2CPZ2}
V=Q_{u}P^{T}_{u}U_{23}(\theta_{u},\delta_{u})\Sigma_{u}^{\dagger}
\Sigma_{d}S_{23}(\theta_{d})P_{d}Q^{\dagger}_{d}\,.
\end{equation}
One can also straightforwardly extract the expression of the CKM mixing matrix for the residual symmetry $Z^{g_{u}}_2\times X_{u}$ and $Z^{g_d}_2$. Before closing this section, we note that the above schemes can be extended to the grand unification theory if both left-handed quarks and leptons could be assigned to the same representation. In the following sections, we assume neutrinos are Majorana particles.

\section{\label{sec:dihedral_group_theory}Dihedral group and CP symmetry}

The dihedral group $D_{n}$ is the symmetry group of an $n$-sided regular polygon for $n>1$.
A regular polygon with $n$ sides has $2n$ different symmetries: $n$ rotational symmetries and $n$ reflection symmetries, therefore the group order of $D_{n}$ is $2n$. All $D_{n}$ are non-ablelian permutation groups  for $n>2$, $D_{1}$ is isomorphic to $Z_{2}$ and $D_{2}$ is isomorphic to $Z_{2}\times Z_{2}$. The group $D_{n}$ is the semidirect product $Z_{n}\rtimes Z_{2}$ of the cyclic groups $Z_{n}$ and $Z_{2}$. The dihedral group can be conveniently defined by two generators $R$ and $S$ which obey the relations,
\begin{equation}
\label{eq:Dn_multi_rules}
R^{n}=S^{2}=(RS)^2=1\,,
\end{equation}
where $R$ refers to rotation and $S$ is the reflection. As a consequence, all the group elements of $D_{n}$ can be expressed as
\begin{equation}
 g=S^{\alpha}R^{\beta}\,
\end{equation}
where $\alpha=0, 1$ and $\beta=0, 1, \dots, n-1$. Then it is straightforward to determine the conjugacy classes of the dihedral group. Depending on whether the group index $n$ is even or odd, the $2n$ group elements of $D_n$ can be classified into three or five types of conjugacy classes.
\begin{itemize}

\item{$n$ is odd}
\begin{equation}
\begin{aligned}
1C_{1}&=\{1\}\,,\\
2C_{m}^{(\rho)}&=\{R^{\rho},R^{-\rho}\}\,,~~\text{with}~~\rho=1,\dots,\frac{n-1}{2}\,,\\
nC_{2}&=\{S,SR,SR^{2},\dots,SR^{n-1}\}\,,
\end{aligned}
\end{equation}
where $m$ is minimal integer such that the identity $m\rho=0 \,(\mathrm{mod}~n)$ is satisfied, and $kC_l$ denotes a conjugacy class of $k$ elements whose order are $l$.

\item{$n$ is even}
\begin{equation}
\begin{aligned}
1C_{1}&=\{1\}\,,\\
1C_{2}&=\{R^{n/2}\}\,,\\
2C_{m}^{(\rho)}&=\{R^{\rho},R^{-\rho}\}\,,~~\text{with}~~\rho=1,\dots,\frac{n-2}{2}\,,\\
\frac{n}{2}C_{2}&=\{S,SR^{2},SR^{4},\dots,SR^{n-4},SR^{n-2}\}\,,\\
\frac{n}{2}C_{2}&=\{SR,SR^{3},\dots,SR^{n-3},SR^{n-1}\}\,,
\end{aligned}
\end{equation}
\end{itemize}
The group structure of $D_n$ is simple, and the subgroups of $D_{n}$ group turn out to be either dihedral or cyclic group. The explicit expressions of all the subgroups are
\begin{eqnarray}
\nonumber Z_{j}&=&<R^{\frac{n}{j}}>\,\quad \text{with}~~j|n\,,\\
\nonumber Z_{2}^{(m)}&=&<SR^{m}>\,\quad \text{with}~~m=0,1,\dots,n-1\,,\\
D_{j}^{(m)}&=&<R^{\frac{n}{j}},SR^{m}>\,\quad\text{with}~~j|n,~ m=0,1,\dots,\frac{n}{j}-1\,.
\end{eqnarray}
Hence the total number of cyclic subgroups generated by certain power of $R$ is equal to the number of positive divisors of $n$, and the total number of dihedral subgroups is the sum of positive divisors of $n$.

The group $D_{n}$ only has real one-dimensional and two-dimensional irreducible representations. The number of irreducible representations is dependent on the parity of the group index $n$.
\begin{itemize}
\item{$n$ is odd}

If the index $n$ is an odd integer, the group $D_{n}$ has two singlet representations $\mathbf{1}_{i}$ and $\frac{n-1}{2}$ doublet representations $\mathbf{2}_{j}$,
where the indices $i$ and $j$ are $i=1, 2$ and $j=1,\dots,\frac{n-1}{2}$.
We observe that the sum of the squares of the dimensions of the irreducible representations is
\begin{equation}
1^2+1^2+2^{2}\times \frac{n-1}{2}=2n\,,
\end{equation}
which is exactly the number of elements in $D_n$ group. In the one-dimensional representations, we have
\begin{equation}
\mathbf{1}_{1}:~R=S=1\,, \quad \mathbf{1}_{2}:~R=1,~S=-1\,.
\end{equation}
For the two-dimensional representations, the generators $R$ and $S$ are represented by
\begin{equation}
\mathbf{2}_{j}:~R=\left(\begin{array}{cc} e^{2\pi i\frac{j}{n}} & 0 \\ 0 & e^{-2\pi i\frac{j}{n}} \end{array} \right)\,, \quad
S=\left( \begin{array}{cc} 0 & 1 \\ 1 & 0 \end{array} \right)\,,
\end{equation}
with $j=1,\dots,\frac{n-1}{2}$.

\item{$n$ is even}

For the case that the index $n$ is an even integer, the group $D_{n}$ has four singlet representations $\mathbf{1}_{i}$ with $i=1,2,3,4$ and $\frac{n}{2}-1$ doublet representations $\mathbf{2}_{j}$ with $j=1,\dots,\frac{n}{2}-1$.
It can be checked that the squared dimensions of the inequivalent irreducible representations add up to the group order as well,
\begin{equation}
1^2+1^2+1^2+1^2+2^{2} \times (\frac{n}{2}-1)=2n\,.
\end{equation}
The generators $R$ and $S$ for the one-dimensional representations are given by
\begin{equation}
\begin{aligned}
&\mathbf{1}_{1}:~R=S=1\,, \qquad~~ ~~~~\mathbf{1}_{2}:~R=1,~S=-1\,,\\
&\mathbf{1}_{3}:~R=-1,~S=1\,, \qquad \mathbf{1}_{4}:~R=S=-1\,.
\end{aligned}
\end{equation}
The explicit forms of these generators in the irreducible two-dimensional representations are
\begin{equation}
\mathbf{2}_{j}:~R=\left(\begin{array}{cc} e^{2\pi i\frac{j}{n}} & 0 \\ 0 & e^{-2\pi i\frac{j}{n}} \end{array} \right)\,, \quad S=\left( \begin{array}{cc} 0 & 1 \\ 1 & 0 \end{array} \right)\,,
\end{equation}
with $j=1,\dots,\frac{n}{2}-1$. Notice that the doublet representation $\mathbf{2}_{j}$ and the complex conjugate $\bar{\mathbf{2}}_{j}$ are unitarily equivalent, and they are related through change of basis, i.e., $R^{*}=URU^{-1}$ and $S^{*}=USU^{-1}$ where $U=\begin{pmatrix} 0 & 1 \\ 1 & 0 \end{pmatrix}$. Hence all the two-dimensional representations of $D_n$ are real representations, although the representation matrix of $R$ is complex in the chosen basis. Moreover, if $a=\left(a_1, a_2\right)^{T}$ is a doublet transforming as $\mathbf{2}_j$, the complex conjugate $\bar{a}=\left(a^{*}_1, a^{*}_2\right)^{T}$
doesn't transform as $\mathbf{2}_j$, but rather $\left(a^{*}_2, a^{*}_1\right)^{T}$ transform as $\mathbf{2}_j$ under $D_n$.

\end{itemize}

In order to consistently combine a flavor symmetry $G_f$ with the CP symmetry, the subsequent action of the CP transformation, an element of the
flavor group and the CP transformation should be equivalent to the action of another element of the flavor group. In other word, the so-called consistency condition has to be fulfilled~\cite{Grimus:1995zi,Feruglio:2012cw,Holthausen:2012dk,Chen:2014tpa}
\begin{equation}
\label{eq:consis_cond}X_{\mathbf{r}}\rho^{*}_{\mathbf{r}}(g)X^{\dagger}_{\mathbf{r}}=\rho_{\mathbf{r}}(g^{\prime}),\quad  g,g^{\prime}\in G_{f}\,,
\end{equation}
where $\rho_{\mathbf{r}}(g)$ is the representation matrix of the element $g$ in the representation $\mathbf{r}$, and $X_{\mathbf{r}}$ is the CP transformation. Here $g$ and $g'$ are generally different group elements,
consequently the CP transformation $X_{\bf r}$ is related to an automorphism which maps $g$ into $g^{\prime}$. In addition, Ref.~\cite{Chen:2014tpa} showed that physical CP transformations should be a class-inverting automorphism of $G_f$, i.e. $g^{-1}$ and $g'$ which is the image of $g$ under the automorphism should be in the same conjugacy class.

We find that the $D_n$ groups really have a class-inverting outer automorphism $\mathfrak{u}$, and its action on the generators is
\begin{equation}
\label{eq:class_inverting_aut_Dn} R\stackrel{\mathfrak{u}}{\longmapsto}R^{-1}\,,\quad S\stackrel{\mathfrak{u}}{\longmapsto}S\,.
\end{equation}
The CP transformation corresponding to $\mathfrak{u}$ is denoted by
$X^0_{\mathbf{r}}$, its concrete form is determined by the following consistency conditions,
\begin{eqnarray}
\nonumber&&X^0_{\mathbf{r}}\rho^{*}_{\mathbf{r}}(R)X^{0\dagger}_{\mathbf{r}}=\rho_{\mathbf{r}}\left(\mathfrak{u}\left(R\right)\right)=\rho_{\mathbf{r}}\left(R^{-1}\right)\,,\\
\label{eq:cons_eq}&&X^0_{\mathbf{r}}\rho^{*}_{\mathbf{r}}(S)X^{0\dagger}_{\mathbf{r}}=\rho_{\mathbf{r}}\left(\mathfrak{u}\left(S\right)\right)=\rho_{\mathbf{r}}\left(S\right)\,.
\end{eqnarray}
In our working basis shown above,
$X^0_{\mathbf{r}}$ is fixed to be a unit matrix up to an overall irrelevant phase,
\begin{equation}
\label{eq:gcp_trans}X^0_{\mathbf{r}}=\mathbb{1}\,.
\end{equation}
Furthermore, including the inner automorphisms, the full set of CP transformations compatible with $D_{n}$ flavor symmetry are
\begin{equation}
\label{eq:GCP_all}X_{\mathbf{r}}=\rho_{\mathbf{r}}(g)X^0_{\mathbf{r}}=\rho_{\mathbf{r}}(g),\quad g\in D_{n}\,.
\end{equation}
Hence the CP transformations compatible with $D_n$ are of the same form as the flavor symmetry transformations in the chosen basis.

\section{\label{sec:Z2CPZ2CP}Mixing patterns from $D_n$ and CP symmetry breaking to two distinct $Z_2\times CP$ subgroups }

In this section, we shall consider the dihedral group $D_n$ as flavor symmetry $G_f$ which is combined with CP symmetry. The three generations of left-handed lepton and quark doublets are assumed to transform as a direct sum of one-dimensional representation $\mathbf{1}_i$ and two-dimensional representation $\mathbf{2}_j$ of $D_n$,
\begin{equation}
\label{eq:assignment}L\sim\left(\mathbf{1}_i, \mathbf{2}_j\right)^{T},\quad Q\sim\left(\mathbf{1}_i, \mathbf{2}_j\right)^{T}\,.
\end{equation}
Notice that usually the three lepton doublets are assigned to an irreducible three-dimensional representation of $G_f$ in order to obtain at least two non-vanishing lepton mixing angles. In the following, we shall show that the singlet plus doublet assignment can also accommodate the experimental data on mixing angles after the CP symmetry is considered. Moreover, one can also assign either the second or the third generation lepton (quark) doublet to a one-dimensional representation of $D_n$ while the remaining two generations transform as a two-dimensional representation of $D_n$. The resulting mixing matrix is related to the corresponding one of the assignment in Eq.~\eqref{eq:assignment} through permutations of rows and columns. Since our approach doesn't make any predictions for the mass spectrum of quarks and leptons, both PMNS mixing matrix and CKM mixing matrix are only determined up to all possible permutations of rows and columns, as shown in Eqs.~(\ref{eq:PMNS_Z2CPZ2CP}, \ref{eq:CKM_Z2CPZ2CP}, \ref{eq:PMNS_Z2CPZ2}, \ref{eq:CKM_Z2CPZ2}). Therefore alternative assignments for the left-handed leptons and quarks mentioned above don't lead to different mixing patterns.

Here we assume that the discrete flavor group $D_{n}$ in combination with CP symmetry is broken down to $Z_{2}\times CP$ in both charged lepton and neutrino sectors, then one entry of the mixing matrix is completely fixed by the residual symmetry. Considering all possible residual symmetries $Z_{2}^{g_{e}}\times X_{e}$ and $Z_{2}^{g_{\nu}}\times X_{\nu}$, we find the fixed element can only be $0,1$ or $\cos\varphi_{1}$, where the value of $\varphi_1$ is determined by the choice of the residual symmetry. Obviously only the mixing pattern with the fixed element $\cos\varphi_{1}$ could be in agreement with experimental data for certain values of $\varphi_{1}$ characterizing the residual symmetry. As a consequence, we find the viable residual symmetries in the lepton sector are $Z_{2}^{g_{e}}=Z_{2}^{SR^{z_{e}}}$, $X_{e}=\{R^{-z_{e}+x},SR^{x}\}$, $Z_{2}^{g_{\nu}}=Z_{2}^{SR^{z_{\nu}}}$ and $X_{\nu}=\{R^{-z_{\nu}+y},SR^{y}\}$, where $z_{e}, z_{\nu}=0,1,\dots,n-1$, $x=y=0$ for odd $n$ and $x, y=0, n/2$ if the group index $n$ is even. Accordingly the lepton mixing matrix reads
\begin{equation}
\label{eq:U_PMNSFin_Z2CPZ2CP}
U_I=\left( \begin{array}{ccc}
\cos\varphi_{1} &~ -c_{\nu}\sin\varphi_{1} &~ s_{\nu}\sin\varphi_{1} \\
c_{e}\sin\varphi_{1} &~ s_{\nu}s_{e}e^{i\varphi_{2}}+c_{\nu}c_{e}\cos\varphi_{1} &~ c_{\nu}s_{e}e^{i\varphi_{2}}-s_{\nu}c_{e}\cos\varphi_{1}\\
-s_{e}\sin\varphi_{1} &~ s_{\nu}c_{e}e^{i\varphi_{2}}-c_{\nu}s_{e}\cos\varphi_{1} &~ c_{\nu}c_{e}e^{i\varphi_{2}}+s_{\nu}s_{e}\cos\varphi_{1} \end{array}\right)\,,
\end{equation}
up to permutations of rows and columns, where
\begin{equation}
s_{e}\equiv\sin\theta_{e}\,,\quad s_{\nu}\equiv \sin\theta_{\nu}\,,\quad c_{e}\equiv\cos\theta_{e}\,,\quad c_{\nu}\equiv \cos\theta_{\nu}\,.
\end{equation}
The parameters $\varphi_{1}$ and $\varphi_{2}$ are discrete group parameters characterizing the residual symmetry, and all possible values of $\varphi_{1}$ and $\varphi_{2}$ are summarized in table~\ref{Tab:values_of_discrete_parameters_lepton_Z2Z2}.
We can see that $\varphi_{1}$ and $\varphi_{2}$ can take the following discrete values
\begin{equation}
\begin{aligned}
&\varphi_{1}~(\text{mod}~2\pi)=0, \frac{1}{n}\pi, \frac{2}{n}\pi, \dots, \frac{2n-1}{n}\pi\,,\\
&\varphi_{2}~(\text{mod}~2\pi)=0, \frac{1}{2}\pi, \pi, \frac{3}{2}\pi\,.
\end{aligned}
\end{equation}
In particular, $\varphi_2$ would be $0$ or $\pi$ if the group index $n$ is odd or $x=y=0, n/2$ for even $n$.
\begin{table}[t!]
\centering
\footnotesize
\begin{tabular}{|c|c|c|c|c|c|c|} \hline \hline
$n$ & $i$ & $x$ & $y$  & $|z_{\nu}-z_{e}|$ &$\varphi_{1}$ & $\varphi_{2}$ \\ \hline
\texttt{odd} & \multirow{5}{*}{$1,2$} & $0$ & $0$& &  \multirow{5}{*}{$j(z_{\nu}-z_{e})\pi/n$} & $0$ \\
\cline{1-1} \cline{3-4} \cline{7-7}
\multirow{12}{*}{\texttt{even}} &  & $0$ & $0$ & & & $0$ \\ \cline{3-4} \cline{7-7}
&  & $n/2$ & $n/2$ &even or odd & & $0$ \\ \cline{3-4} \cline{7-7}
&  & $0$ & $n/2$ & & & $-j\pi/2$ \\ \cline{3-4} \cline{7-7}
&  & $n/2$ & $0$ & & & $j\pi/2$  \\ \cline{2-7}
& \multirow{8}{*}{$3,4$} & \multirow{2}{*}{$0$} & \multirow{2}{*}{$0$} &even &$j(z_{\nu}-z_{e})\pi/n$ & $(z_{\nu}-z_{e})\pi/2$ \\ \cline{5-7}
&  &  &  & odd &$j(z_{\nu}-z_{e})\pi/n+\pi/2$ & $(z_{\nu}-z_{e})\pi/2+\pi/2$ \\  \cline{3-7}
&  & \multirow{2}{*}{$n/2$} & \multirow{2}{*}{$n/2$} & even &$j(z_{\nu}-z_{e})\pi/n$ & $(z_{\nu}-z_{e})\pi/2$ \\ \cline{5-7}
&  &  &  & odd &$j(z_{\nu}-z_{e})\pi/n+\pi/2$ & $(z_{\nu}-z_{e})\pi/2+\pi/2$ \\ \cline{3-7}
&  & \multirow{2}{*}{$0$} & \multirow{2}{*}{$n/2$} & even &$j(z_{\nu}-z_{e})\pi/n$ & $(z_{\nu}-z_{e})\pi/2-(2j+n)\pi/4$ \\ \cline{5-7}
&  &  &  & odd &$j(z_{\nu}-z_{e})\pi/n+\pi/2$ & $(z_{\nu}-z_{e})\pi/2-(2j+n)\pi/4+\pi/2$ \\ \cline{3-7}
&  & \multirow{2}{*}{$n/2$} & \multirow{2}{*}{$0$} & even &$j(z_{\nu}-z_{e})\pi/n$ & $(z_{\nu}-z_{e})\pi/2+(2j+n)\pi/4$ \\ \cline{5-7}
&  &  &  & odd &$j(z_{\nu}-z_{e})\pi/n+\pi/2$ & $(z_{\nu}-z_{e})\pi/2+(2j+n)\pi/4+\pi/2$ \\ \hline \hline
\end{tabular}
\caption{\label{Tab:values_of_discrete_parameters_lepton_Z2Z2}The values of the parameters $\varphi_{1}$ and $\varphi_{2}$ for the residual symmetries  $Z_{2}^{g_{e}}=Z_{2}^{SR^{z_{e}}}$, $X_{e}=\{R^{-z_{e}+x},SR^{x}\}$, $Z_{2}^{g_{\nu}}=Z_{2}^{SR^{z_{\nu}}}$ and $X_{\nu}=\{R^{-z_{\nu}+y},SR^{y}\}$, where $i$ and $j$ are the indces of the representations $\mathbf{1}_{i}$ and $\mathbf{2}_j$.}
\end{table}
It is easy to check that the mixing matrix in Eq.~\eqref{eq:U_PMNSFin_Z2CPZ2CP} has the following symmetry properties,
\begin{eqnarray}
\nonumber
U_{I}(\varphi_{1},\varphi_{2}, \pi+\theta_{e},\theta_{\nu})&=&\text{diag}(1,-1,-1)U_{I}(\varphi_{1},\varphi_{2},\theta_{e},\theta_{\nu})\,,\\
\nonumber
U_{I}(\varphi_{1},\varphi_{2},\pi-\theta_{e},\theta_{\nu})&=&\text{diag}(1,-1,1)U_{I}(\varphi_{1},\varphi_{2},\theta_{e},\pi-\theta_{\nu})\text{diag}(1,-1,1)\,,\\
\nonumber
U_{I}(\varphi_{1},\varphi_{2},\theta_{e}, \pi+\theta_{\nu})&=&U_{I}(\varphi_{1},\varphi_{2},\theta_{e},\theta_{\nu})\text{diag}(1,-1,-1)\,,\\
\nonumber
U_{I}(\pi+\varphi_{1},\varphi_{2},\theta_{e},\theta_{\nu})&=&\text{diag}(-1,1,-1)U_{I}(\varphi_{1},\varphi_{2},\pi-\theta_{e},\theta_{\nu})\,,\\
\nonumber
U_{I}(\pi-\varphi_{1},\varphi_{2},\theta_{e},\theta_{\nu})&=&\text{diag}(-1,1,1)U_{I}(\varphi_{1},\varphi_{2},\theta_{e},\pi-\theta_{\nu})\text{diag}(1,1,-1)\,,\\
\nonumber U_{I}(\varphi_{1}, \pi+\varphi_{2},\theta_{e},\theta_{\nu})&=&U_{I}(\varphi_{1},\varphi_{2},\theta_{e},\pi-\theta_{\nu})\text{diag}(1,-1,1)\,,\\
\label{eq:PMNS_relations_I}
U_{I}(\varphi_{1},\pi-\varphi_{2},\theta_{e},\theta_{\nu})&=&U_{I}^{*}(\varphi_{1},\varphi_{2},\theta_{e},\pi-\theta_{\nu})\text{diag}(1,-1,1)\,,
\end{eqnarray}
where the above diagonal matrices can be absorbed into the lepton fields. Therefore the parameters $\varphi_1$ and $\varphi_{2}$ can be limited in the ranges $0\leq\varphi_1\leq\pi/2$ and $0\leq\varphi_{2}<\pi$ respectively, and the values of $\varphi_2$ are $0$ and $\pi/2$ in the fundamental region. In the case of $\varphi_2=0$, the mixing matrix in Eq.~\eqref{eq:U_PMNSFin_Z2CPZ2CP} is real such that all CP phases are conserved. The evidence of CP violation in neutrino oscillation has been reported by T2K~\cite{Abe:2017vif} and NO$\nu$A~\cite{NOvA:2018gge} and nontrivial $\delta_{CP}$ is also preferred by global data analysis~\cite{Esteban:2018azc}. Hence we shall focus on $\varphi_2=\pi/2$ in the following.

Including the row and column permutations encoded in $P_e$ and $P_{\nu}$, we find that the 36 possible permutations of rows and columns give rise to nine independent mixing patterns,
\begin{equation}
\label{eq:PMNS_caseI}
\begin{aligned}
&U_{I,1}=U_{I},\qquad\qquad~~ U_{I,2}=U_{I}P_{12},\qquad\qquad~~ U_{I,3}=U_{I}P_{13}\,, \\
&U_{I,4}=P_{12}U_{I}, \qquad~~~ U_{I,5}=P_{12}U_{I}P_{12},\qquad~~~~ U_{I,6}=P_{12}U_{I}P_{13}\,, \\
&U_{I,7}=P_{23}P_{12}U_{I}, \quad~~ U_{I,8}=P_{23}P_{12}U_{I}P_{12}, \quad~~ U_{I,9}=P_{23}P_{12}U_{I}P_{13}\,.
\end{aligned}
\end{equation}
That is to say, the fixed element $\cos\varphi_1$ can be in any position of the mixing matrix. For each mixing pattern, we can straightforwardly extract the expressions of the mixing angles $\sin^{2}\theta_{13}$, $\sin^{2}\theta_{12}$, $\sin^{2}\theta_{23}$ and the CP invariants $J_{CP}$, $I_{1}$, and $I_{2}$, as shown in table~\ref{tab:mixing_parameter_I}. We see that the mixing matrices $U_{I,7}$, $U_{I,8}$ and $U_{I,9}$ can be obtained from $U_{I,4}$, $U_{I,5}$ and $U_{I,6}$ through an exchange of the
second and third rows. Therefore they lead to the same solar mixing angle, reactor mixing angle and Majorana CP phases while atmospheric angle changes from $\theta_{23}$ to $\pi/2-\theta_{23}$ and the Dirac CP phase changes from $\delta_{CP}$ to $\pi+\delta_{CP}$. For the mixing pattern $U_{I, 4}$ and the concerned value $\varphi_2=\pi/2$,
the smallness of $\theta_{13}$ requires
\begin{equation}
\theta_{\nu}\simeq \theta_{e}\simeq0, \qquad \mathrm{or} \qquad \theta_{\nu}\simeq\theta_{e}\simeq\frac{\pi}{2}\,.
\end{equation}
Consequently we have
\begin{equation}
\sin^2\theta_{23}\simeq0,\qquad \mathrm{or} \qquad  \sin^2\theta_{12}\simeq1\,,
\end{equation}
which is obviously not compatible with the measured values of $\theta_{12}$ and $\theta_{23}$~\cite{Esteban:2018azc,deSalas:2017kay,Capozzi:2018ubv}. Therefore this mixing pattern can not accommodate the experimental data. Analogously the mixing patterns $U_{I, 5}$, $U_{I, 7}$ and $U_{I, 8}$ are not viable as well. Furthermore, all the mixing angles and CP phases are expressed in terms of two continuous parameters $\theta_{e}$ and $\theta_{\nu}$ as well as the discrete parameter $\varphi_{1}$ such that strong correlations among the mixing parameters are expected. Eliminating the rotation angles $\theta_{e}$ and $\theta_{\nu}$, we can find the the following sum rules among mixing angles and Dirac CP violation phase,
\begin{table}[t!]
\centering
\small
{
\begin{tabular}{|c|c|c|c|c|c|c|c|c|c|}
\hline \hline
 \texttt{PMNS} & $\varphi_{1}/\pi$ & $n_{\text{min}}$ & \texttt{Mixing Parameters}  \\
 &  &&      \\ [-0.16in]\hline

 &  &&  \\ [-0.16in]
\multirow{4}{*}[-5pt]{$U_{I,1}$}  &\multirow{4}{*}[-5pt]{$[0.181,0.206]$}  & \multirow{4}{*}[-5pt]{$10$} &  $\sin^2\theta_{13}=s_{\nu}^2 \sin ^2\varphi_{1} $ \\
  &  &&     \\ [-0.16in] \cline{4-4}
     &  &&   \\ [-0.16in]
 & &  &   $\sin^2\theta_{12}=\frac{c_{\nu}^2 \sin ^2\varphi_{1}}{1-s_{\nu}^2 \sin ^2\varphi_{1}}$ \\
  &  &&     \\ [-0.16in] \cline{4-4}
   &&    &    \\ [-0.16in]
 &  &&    $\sin^2\theta_{23}=\frac{s_{\nu}^2 c_{e}^2 \cos ^2\varphi_{1}+s_{e}^2 c_{\nu}^2- \mathcal{F} \cos \varphi_{2}}{1-s_{\nu}^2 \sin ^2\varphi_{1}}$ \\
 &  &&      \\ [-0.16in]\cline{4-4}

 &  &&  \\ [-0.16in]

   &  &&  $I_1=I_2=0$   \\
   &  &&      \\ [-0.16in] \hline

   &  &&     \\ [-0.16in]
\multirow{4}{*}[-5pt]{$U_{I,2}$} & \multirow{4}{*}[-5pt]{$[0.301,0.327]$}   & \multirow{4}{*}[-5pt]{$16$} &   $\sin^2\theta_{13}=s_{\nu}^2 \sin ^2\varphi_{1}$ \\
  &   &&     \\ [-0.16in] \cline{4-4}
    & &    &    \\ [-0.16in]
 &  &  &    $\sin^2\theta_{12}=\frac{\cos ^2\varphi_{1}}{1-s_{\nu}^2 \sin ^2\varphi_{1}}$ \\
  &  &&     \\ [-0.16in] \cline{4-4}
    &   &&    \\ [-0.16in]
 &  &&    $\sin^2\theta_{23}=\frac{s_{\nu}^2 c_{e}^2 \cos ^2\varphi_{1}+s_{e}^2 c_{\nu}^2- \mathcal{F} \cos \varphi_{2}}{1-s_{\nu}^2 \sin ^2\varphi_{1}}$ \\
 &  &&      \\ [-0.16in]\cline{4-4}

 &  &&  \\ [-0.16in]

   &  &&  $I_1=I_2=0$   \\

  &  &&      \\ [-0.16in] \hline

  &  &&     \\ [-0.16in]
\multirow{4}{*}[-1pt]{$U_{I,3}$} & \multirow{4}{*}[-1pt]{$[0.450,0.454]$} & \multirow{4}{*}[-1pt]{$30$} &    $\sin^2\theta_{13}=\cos ^2\varphi_{1}$ \\
  &  &&      \\ [-0.16in] \cline{4-4}
    & &    &    \\ [-0.16in]
 &  &  &     $\sin^2\theta_{12}=c_{\nu} ^2$ \\
  &  &&      \\ [-0.16in] \cline{4-4}
    &  &&     \\ [-0.16in]
 &  &&    $\sin^2\theta_{23}= c_{e}^2  $ \\
 &  &&      \\ [-0.16in]\cline{4-4}

 &  &&  \\ [-0.16in]

   &  &&  $I_1=I_2=0$   \\

  &  &&      \\ [-0.16in] \hline

  &  &&     \\ [-0.16in]
\multirow{6}{*}{$U_{I,4}$} & \multirow{6}{*}{$[0.339,0.424]$}  & \multirow{6}{*}{$-$} &    $\sin^2\theta_{13}=s_{\nu}^2 c_{e}^2 \cos ^2\varphi_{1}+s_{e}^2 c_{\nu}^2- \mathcal{F} \cos \varphi_{2} $ \\
  &   &&     \\ [-0.16in] \cline{4-4}
     &&   &    \\ [-0.16in]
 &  &&    $\sin^2\theta_{12}=1-\frac{ c_{e}^2  \sin^2 \varphi_{1}}{1-s_{\nu}^2 c_{e}^2 \cos ^2\varphi_{1}-s_{e}^2 c_{\nu}^2+\mathcal{F} \cos \varphi_{2}}$ \\
  &   &&     \\ [-0.16in] \cline{4-4}
    &   &&    \\ [-0.16in]
 &  &  &    $\sin^2\theta_{23}=\frac{ s_{\nu}^2\sin^2 \varphi_{1} }{1-s_{\nu}^2 c_{e}^2 \cos ^2\varphi_{1}-s_{e}^2 c_{\nu}^2+\mathcal{F} \cos \varphi_{2}}$ \\
  &   &&     \\ [-0.16in] \cline{4-4}

  &   &&    \\ [-0.16in]
 &   &&   $|I_{1}|= |2 s_{e} s_{\nu} c_{e}^2 \sin ^2\varphi_{1} \sin \varphi_{2} \left(s_{e} s_{\nu} \cos \varphi_{2}+c_{e} c_{\nu} \cos \varphi_{1}\right)|$ \\
  &    &&    \\ [-0.16in] \cline{4-4}
    &   &&    \\ [-0.16in]
 &  &&    $|I_{2}|= |2 s_{e} c_{e}^2 c_{\nu} \sin ^2\varphi_{1} \sin \varphi_{2} \left(s_{e} c_{\nu} \cos \varphi_{2}-s_{\nu} c_{e} \cos \varphi_{1}\right)|$ \\
  &   &&     \\ [-0.16in] \hline

   &   &&    \\ [-0.16in]
\multirow{6}{*}{$U_{I,5}$} & \multirow{6}{*}{$[0.266,0.349]$}  & \multirow{6}{*}{$-$} &     $\sin^2\theta_{13}=s_{\nu}^2 c_{e}^2 \cos ^2\varphi_{1}+s_{e}^2 c_{\nu}^2- \mathcal{F} \cos \varphi_{2}  $ \\
  &   &&     \\ [-0.16in] \cline{4-4}
     &  &&     \\ [-0.16in]
 &  &&    $\sin^2\theta_{12}=\frac{c_{e}^2 \sin ^2\varphi_{1}}{1-s_{\nu}^2 c_{e}^2 \cos ^2\varphi_{1}-s_{e}^2 c_{\nu}^2+\mathcal{F} \cos \varphi_{2}}$ \\
  &  &&      \\ [-0.16in] \cline{4-4}
    &   &&    \\ [-0.16in]
 &  &&    $\sin^2\theta_{23}=\frac{s_{\nu}^2\sin^2 \varphi_{1}  }{1-s_{\nu}^2 c_{e}^2 \cos ^2\varphi_{1}-s_{e}^2 c_{\nu}^2+\mathcal{F} \cos \varphi_{2}}$ \\
  &  &&      \\ [-0.16in] \cline{4-4}

  &   &&    \\ [-0.16in]
 &  &&    $|I_{1}|=|2 s_{e} s_{\nu} c_{e}^2 \sin ^2\varphi_{1} \sin \varphi_{2} \left(s_{e} s_{\nu} \cos \varphi_{2}+c_{e} c_{\nu} \cos \varphi_{1}\right)|$ \\
  &  &&      \\ [-0.16in] \cline{4-4}
    &   &&    \\ [-0.16in]
 &  &&  $|I_{2}|= |\mathcal{F} \sin \varphi_{2}  (s_{e} ^2-c_{e}^2 \cos ^2\varphi_{1}-2c_{e}s_{e} \cot 2\theta_{\nu} \cos \varphi_{1} \cos \varphi_{2})|$\\

  &   &&     \\ [-0.16in] \hline

   &  &&     \\ [-0.16in]
\multirow{6}{*}{$U_{I,6}$} & \multirow{6}{*}{$[0.215,0.276]$}  & \multirow{6}{*}{$4$} &     $\sin^2\theta_{13}=c_{e}^2 \sin ^2\varphi_{1}$ \\
  &   &&     \\ [-0.16in] \cline{4-4}
     &  &&     \\ [-0.16in]
 &    &&  $\sin^2\theta_{12}=\frac{s_{e}^2 s_{\nu}^2+c_{e}^2 c_{\nu}^2 \cos ^2\varphi_{1}+\mathcal{F} \cos \varphi_{2}}{1-c_{e}^2 \sin ^2\varphi_{1}}$ \\
  &   &&     \\ [-0.16in] \cline{4-4}
    &  &&     \\ [-0.16in]
 &  &&    $\sin^2\theta_{23}=\frac{\cos ^2\varphi_{1}}{1-c_{e}^2\sin ^2\varphi_{1}}$ \\
  &   &&     \\ [-0.16in] \cline{4-4}

  &  &&     \\ [-0.16in]
 &  &&    $|I_{1}|=|\mathcal{F} \sin \varphi_{2}   (2c_{e}s_{e} \cot2 \theta_{\nu} \cos \varphi_{1} \cos \varphi_{2}+  c_{e}^2 \cos ^2\varphi_{1}-s_{e}^2)|$ \\
  &  &&      \\ [-0.16in] \cline{4-4}
    &&   &    \\ [-0.16in]
 &  &&    $|I_{2}|=| 2 s_{e} c_{e}^2 c_{\nu} \sin ^2\varphi_{1} \sin \varphi_{2} \left(s_{\nu} c_{e} \cos \varphi_{1}-s_{e} c_{\nu} \cos \varphi_{2}\right)|$ \\ \hline \hline

\end{tabular}}
\caption{\label{tab:mixing_parameter_I}The predictions of the lepton mixing parameters for the mixing patterns in Eq.~\eqref{eq:PMNS_caseI}.
The absolute value of the Jarlskog invariant $J_{CP}$ is the same for all the nine mixing patterns, i.e. $|J_{CP}|=\frac{1}{2}|\sin2\theta_{e}\sin2\theta_{\nu} \sin^2\varphi_{1}\cos\varphi_{1}\sin\varphi_{2}|$. The parameter $\mathcal{F}$ is defined as $\mathcal{F}=\frac{1}{2}\sin 2\theta_{e} \sin 2\theta_{\nu}\cos\varphi_{1}$. The allowed region of $\varphi_1$ in the second column is obtained by requiring the fixed element $\cos\varphi_1$ is in the experimentally preferred $3\sigma$ range~\cite{Esteban:2018azc}. The notation $n_{\text{min}}$ denotes the minimal value of $n$ which can accommodate the measured values of the lepton mixing angles for $\varphi_2=\pi/2$. Notice that only the value of $\varphi_2=\pi/2$ in its fundamental interval can generate a non-trivial Dirac CP phase in the scenario of section~\ref{sec:Z2CPZ2CP}.  }
\end{table}

\begin{subequations}
\begin{eqnarray}
\label{eq:correlation_Ia}&&U_{I,1}:~\cos^2\theta_{13}\cos^{2}\theta_{12}=\cos^2\varphi_{1}\,,~\sin^{2}\theta_{23}=\frac{1}{2}\pm \sqrt{1-4x^{2}}(\frac{1}{2}-\cot^{2}\varphi_{1}\tan^{2}\theta_{13})\,,\\
\label{eq:correlation_Ib}&&U_{I,2}:~\cos^{2}\theta_{13}\sin^2\theta_{12}=\cos^2\varphi_{1}\,,~\sin^{2}\theta_{23}=\frac{1}{2}\pm \sqrt{1-4x^{2}}(\frac{1}{2}-\cot^{2}\varphi_{1}\tan^{2}\theta_{13})\,,\\
\label{eq:correlation_Ic}&&U_{I,6}:~\cos^{2}\theta_{13}\sin^2\theta_{23}=\cos^2\varphi_{1}\,,~\sin^{2}\theta_{12}=\frac{1}{2}\pm \sqrt{1-4x^{2}}(\frac{1}{2}-\cot^{2}\varphi_{1}\tan^{2}\theta_{13})\,,\\
\label{eq:correlation_Id}&&U_{I,9}:~\cos^{2}\theta_{13}\cos^2\theta_{23}=\cos^2\varphi_{1}\,,~\sin^{2}\theta_{12}=\frac{1}{2}\pm \sqrt{1-4x^{2}}(\frac{1}{2}-\cot^{2}\varphi_{1}\tan^{2}\theta_{13})\,,
\end{eqnarray}
\end{subequations}
where ``+'' is for $\theta_{e}\in\left(\pi/4, 3\pi/4\right)$ and ``$-$'' for $\theta_{e}\in[0,\pi/4]\cup[3\pi/4, \pi]$, and the parameter $x$ is defined as
\begin{equation}\label{eq:par_x_UI}
x=\frac{J_{CP}}{\sin\theta_{13}\cos\varphi_{1}\sqrt{\sin^2\varphi_{1}-\sin^2\theta_{13}}}\,.
\end{equation}
The current and upcoming neutrino experiments will be able to significantly reduce the experimental errors on $\theta_{12}$ and $\theta_{23}$, and the next generation long-baseline experiments
are expected to considerably improve the sensitivity to the Dirac phase $\delta_{CP}$ if running in both the neutrino and the anti-neutrino modes. Thus the above sum rules could be tested and possibly distinguished from each other in future, or to be ruled out entirely.

\begin{figure}[t!]
\centering
\begin{tabular}{cc}
\includegraphics[width=0.48\linewidth]{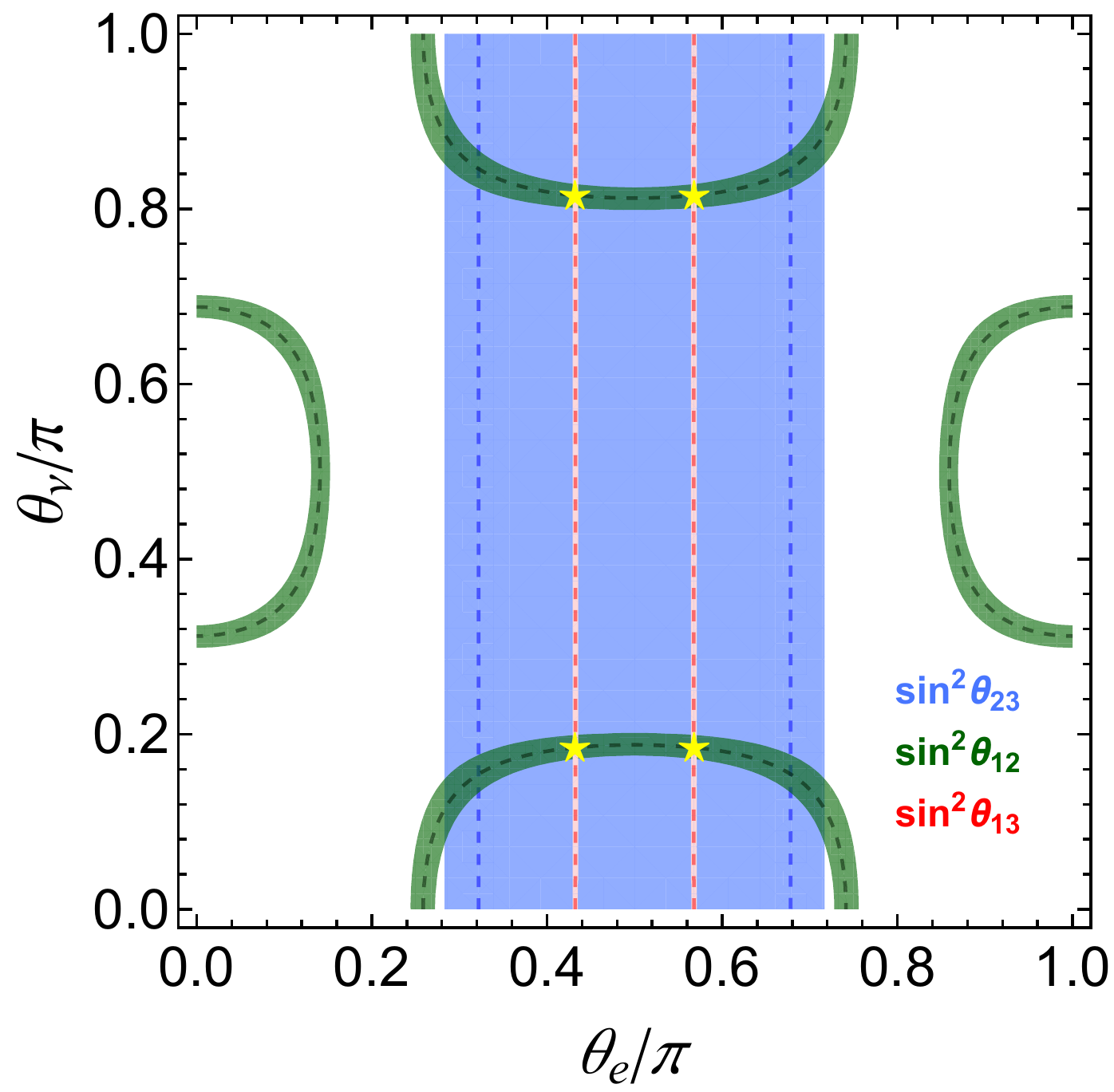}
\includegraphics[width=0.48\linewidth]{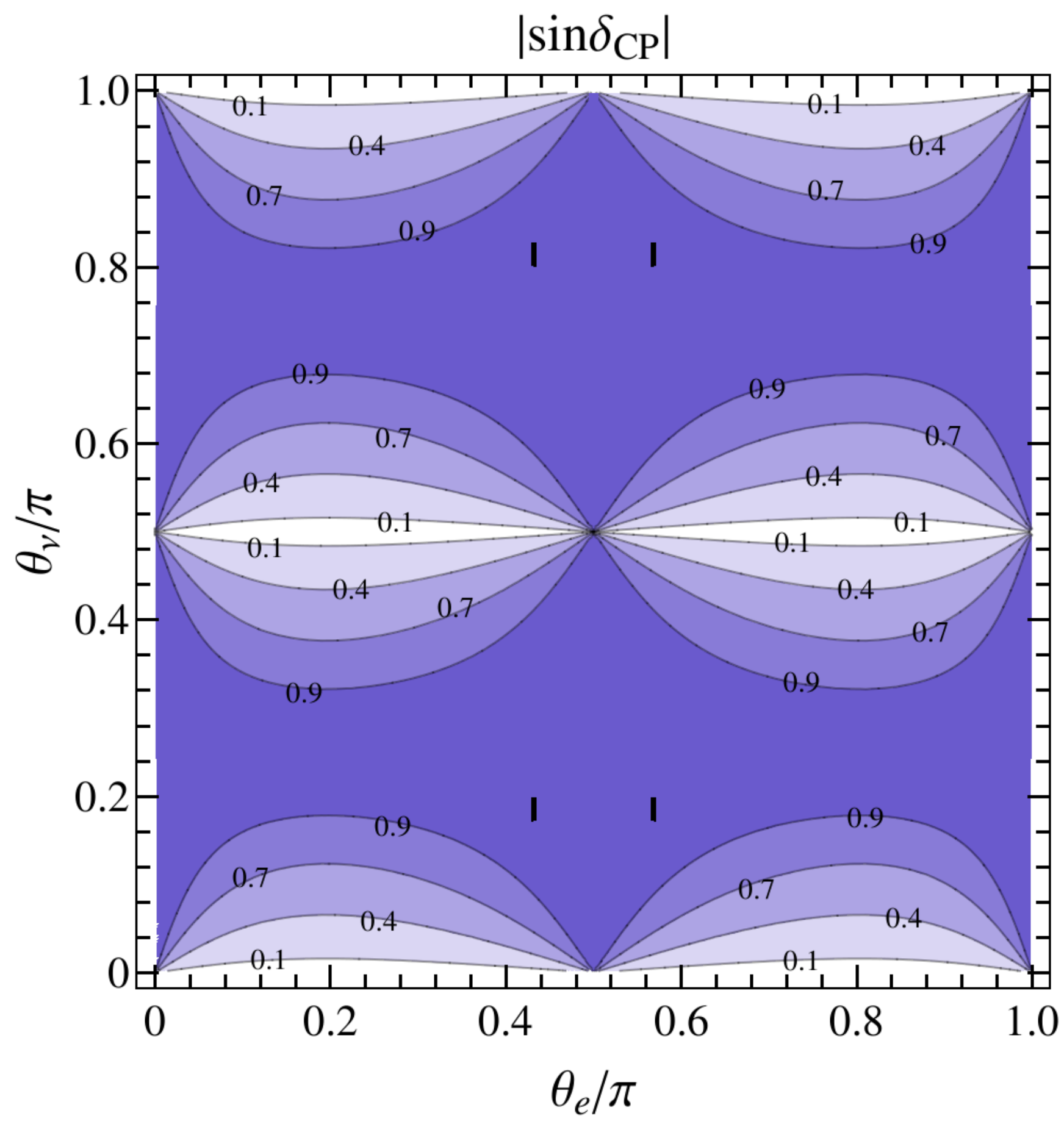}\\
\includegraphics[width=0.48\linewidth]{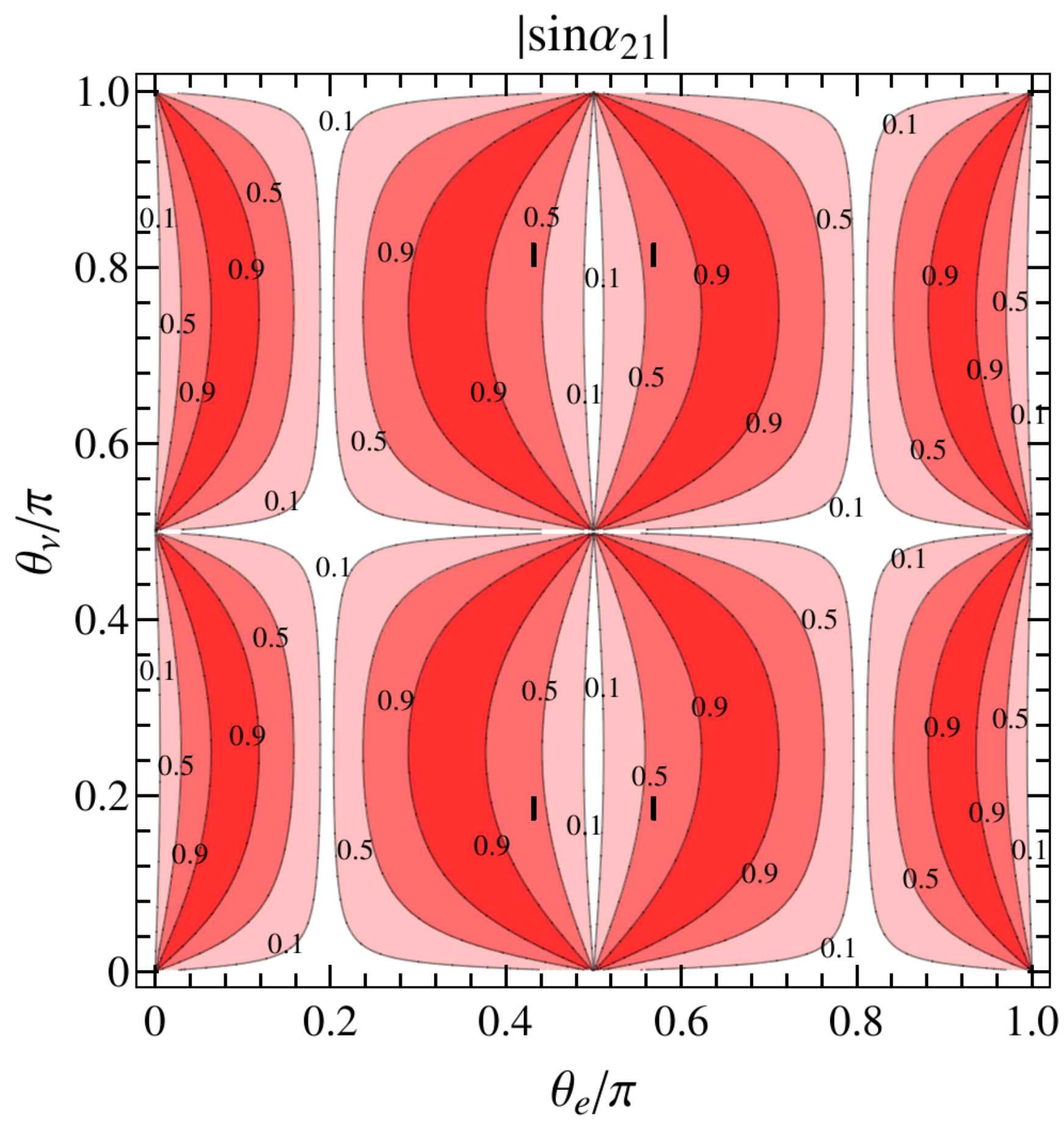}
\includegraphics[width=0.48\linewidth]{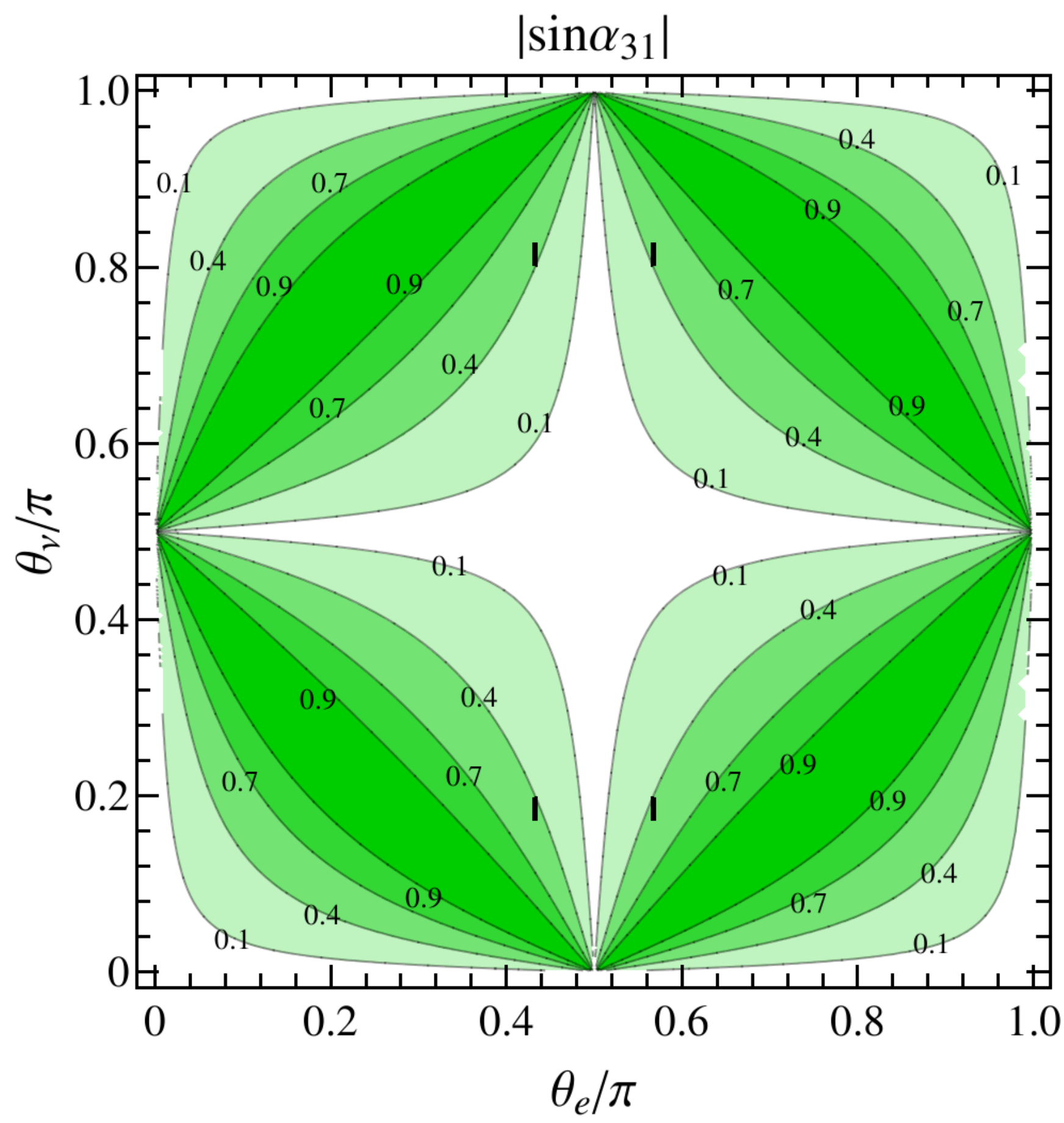}
\end{tabular}
\caption{\label{fig_contour:case_I} The contour plots of $\sin^2\theta_{ij}$ and the CP violation phases $|\sin\delta_{CP}|$, $|\sin\alpha_{21}|$ and $|\sin\alpha_{31}|$ in the plane $\theta_{\nu}$ versus $\theta_{e}$ for the mixing pattern $U_{I,6}$ with $\varphi_{1}=\pi/4$ and $\varphi_{2}=\pi/2$. The upper-left panel is the contour plots of $\sin^2\theta_{ij}$. The red, green and blue areas denote the $3\sigma$ contour regions of $\sin^2\theta_{13}$, $\sin^2\theta_{12}$ and $\sin^2\theta_{23}$ respectively. The dashed contour lines stand for the corresponding experimental best fit values. The $3\sigma$ ranges and the best fit values of the mixing angles are taken from~\cite{Esteban:2018azc}. The best fitting values of $\theta_{e, \nu}$ are indicated with yellow pentagrams. The black areas represent the regions for which all the three lepton mixing angles are compatible with experimental data at
$3\sigma$ level.}
\end{figure}

It is notable that the experimental data on lepton mixing angles can be explained by small dihedral group $D_4$ which is the symmetry group of a square. For instance, if the residual symmetry is specified by $x=2$, $y=0$, $z_{\nu}=1$, $z_{e}=0$ and the three left-handed lepton doublets are assigned to $\mathbf{1}_1\oplus\mathbf{2}_1$, we have $\varphi_1=\pi/4$ and $\varphi_2=\pi/2$, the mixing angles for the mixing matrices $U_{I,6}$ and $U_{I,9}$ can be quite close to their best fit values for certain choices of parameters $\theta_{\nu}$ and $\theta_{e}$. As a measure for the goodness of fit, we perform a global fit using the $\chi^2$ function which is defined in the usual way\footnote{The information of $\delta_{CP}$ is not included in the $\chi^2$ function because it is measured with large uncertainties at present and the indication of its preferred value from global data analyses is rather weak~\cite{Esteban:2018azc}.}. The numerical results are listed in table~\ref{tab:bf_UI_n=4}. We can see that the deviation of $\theta_{23}$ from maximal value can be accommodated and the Dirac phase $\delta_{CP}$ could be around $1.5\pi$. Moreover, we display the contour regions for $\sin^2\theta_{ij}$, $|\sin\delta_{CP}|$, $|\sin\alpha_{21}|$ and $|\sin\alpha_{31}|$ in figure~\ref{fig_contour:case_I}. As one can clearly see, the rotation angles $\theta_e$ and $\theta_{\nu}$ are strongly constrained to accommodate the three lepton mixing angles $\theta_{ij}$ within the
experimentally preferred $3\sigma$ intervals (black areas in the figure). Therefore the allowed ranges of the mixing angles and CP phases should be rather narrow around the numerical values in table~\ref{tab:bf_UI_n=4} and the present approach is very predictive.

\begin{figure}[t!]
\centering
\begin{tabular}{cc}
\includegraphics[width=0.55\linewidth]{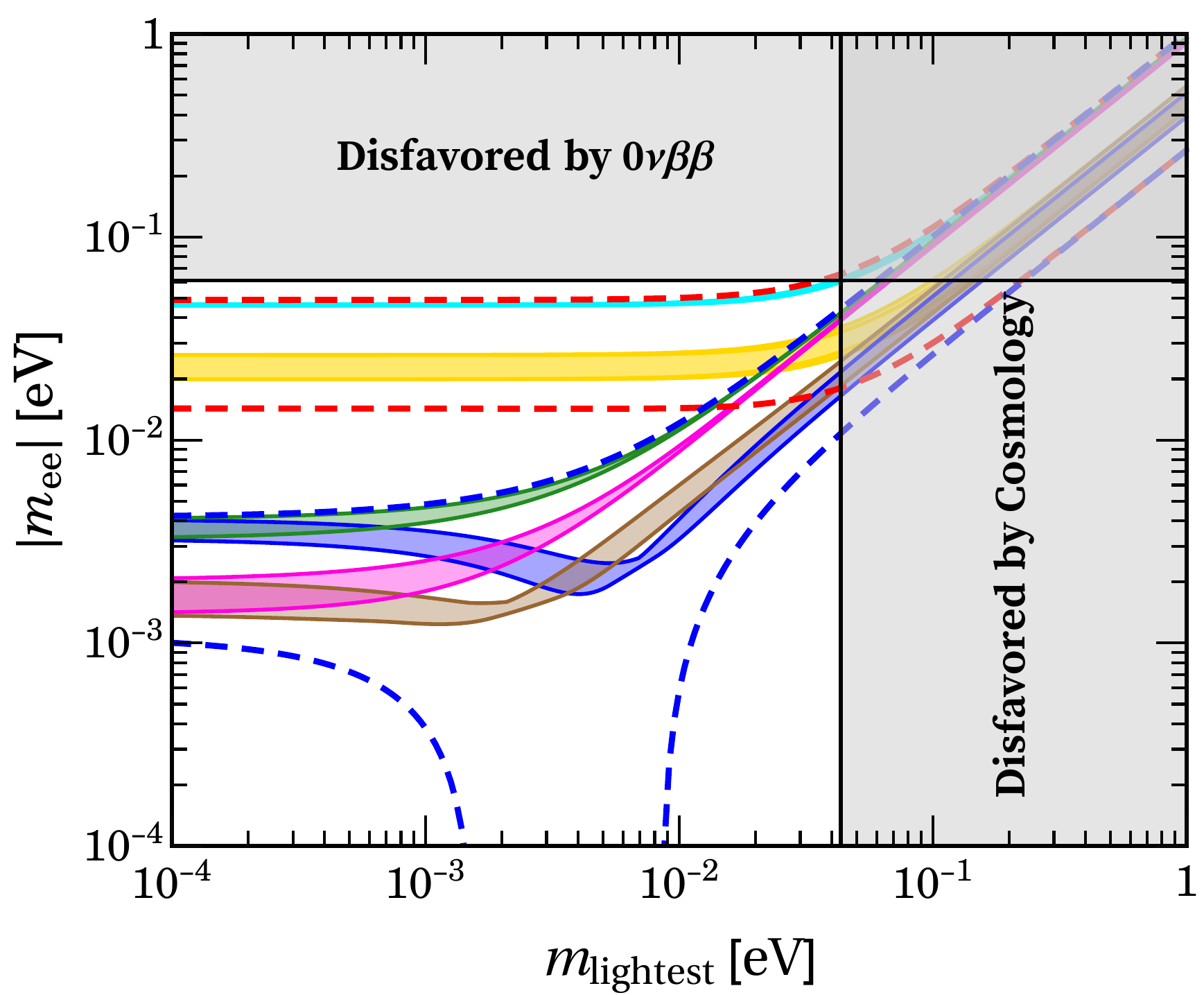}&~~~
\multirow{1}[30]{0.42\linewidth}[6.1cm]{\vskip0.01cm\includegraphics[width=0.88\linewidth]{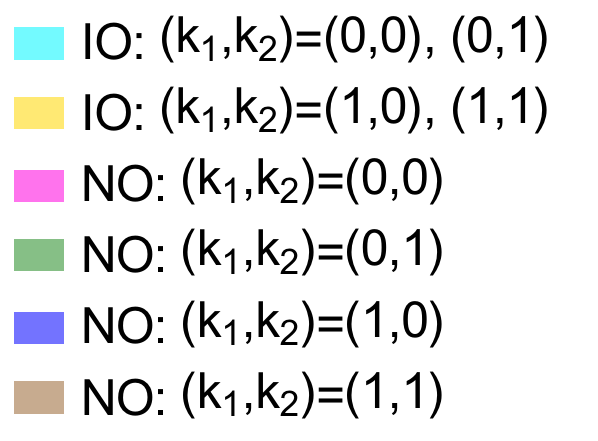}}
\end{tabular}
\caption{\label{fig:mee_I} The possible values of the effective Majorana mass $m_{ee}$ as a function of the lightest neutrino mass $m_{\rm lightest}$ for the mixing pattern $U_{I,6}$ with $n=4$ and $\varphi_{1}=\pi/4$, $\varphi_{2}=\pi/2$.  The red (blue) dashed lines indicate the most general allowed regions for IO (NO) neutrino mass spectrum obtained by varying the mixing parameters over their $3\sigma$ ranges~\cite{Esteban:2018azc}. The present most stringent upper limits $|m_{ee}|<0.120$ eV from EXO-200~\cite{Auger:2012ar, Albert:2014awa} and KamLAND-ZEN~\cite{Gando:2012zm} is represented by horizontal grey band. The vertical grey exclusion band denotes the current bound coming from the cosmological data of $\sum m_i<0.130$ eV at $95\%$ confidence level obtained by the Planck collaboration~\cite{Aghanim:2018eyx}. }
\end{figure}

The neutrinoless double beta ($0\nu\beta\beta$) decay is the unique probe for the Majorana nature of neutrinos, and it explicitly depends on the values of the Majorana CP violation phases. The $0\nu\beta\beta$ decay experiments can provide valuable information on the neutrino mass spectrum and constrain the Majorana phases. The $0\nu\beta\beta$ decay rate is proportional to the effective Majorana mass $|m_{ee}|$ which is the (11) element of the neutrino mass matrix in the charged lepton diagonal basis,
\begin{equation}
|m_{ee}|=|m_1\cos^2\theta_{12}\cos^2\theta_{13}+m_2\sin^2\theta_{12}\cos^2\theta_{13}e^{i\alpha_{21}}+m_3\sin^2\theta_{13}e^{i(\alpha_{31}-2\delta_{CP})}|\,.
\end{equation}
For the mixing pattern $U_{I, 6}$ with $\varphi_1=\pi/4$ and $\varphi_2=\pi/2$, we show the prediction of the effective Majorana mass $|m_{ee}|$ as a function of the lightest neutrino mass $m_{\text{lightest}}$ in figure~\ref{fig:mee_I}. If the neutrino mass spectrum is IO, the effective mass $|m_{ee}|$ is around 0.046 eV or in the narrow interval $[0.020 \mathrm{eV}, 0.026 \mathrm{eV}]$. Since the next generation $0\nu\beta\beta$ decay experiments will be able to explore the whole region of the IO parameter space such that these predictions could be tested. In the case of NO spectrum, the effective Majorana mass has a lower limit $|m_{ee}|\geq0.0011$ eV.

\begin{table}[t!]
\centering
\footnotesize
\renewcommand{\tabcolsep}{1.6mm}
\begin{tabular}{|c|c|c|c|c|c|c|c|c|c|c|c|c|}
\hline \hline
& $\varphi_1$ & $\varphi_2$ & MO& $\theta^{\text{bf}}_{e}/\pi$ & $\theta^{\text{bf}}_{\nu}/\pi$ & $\chi^2_{\text{min}}$   & $\sin^2\theta_{13}$ & $\sin^2\theta_{12}$ & $\sin^2\theta_{23}$  & $\delta_{CP}/\pi$ & $\alpha_{21}/\pi~(\text{mod}~1)$ & $\alpha_{31}/\pi~(\text{mod}~1)$  \\ \hline
\multirow{8}{*}{$U_{I, 6}$}&\multirow{16}{*}{$\frac{\pi}{4}$} & \multirow{16}{*}{$\frac{\pi}{2}$} & \multirow{4}{*}{NO}
& $0.432$ & $0.185$ & \multirow{4}{*}{$13.781$}  & \multirow{4}{*}{$0.0224$} & \multirow{4}{*}{$0.310$} & \multirow{4}{*}{$0.511$}   & \multirow{2}{*}{$1.459$} & \multirow{2}{*}{$0.790$} & \multirow{2}{*}{$0.854$}  \\ \cline{5-6}
&&&& $0.568$ & $0.815$ &   &  &  &    &  &  &   \\ \cline{5-6} \cline{11-13}
&&&& $0.432$ & $0.815$ &   &  &  &    & \multirow{2}{*}{$0.541$} & \multirow{2}{*}{$0.210$} & \multirow{2}{*}{$0.146$}  \\ \cline{5-6}
&&&& $0.568$ & $0.185$ &   &  &  &    &  &  &   \\  \cline{4-13}

&&&\multirow{4}{*}{IO} & $0.432$ & $0.185$ & \multirow{4}{*}{$15.301$} & \multirow{4}{*}{$0.0227$} & \multirow{4}{*}{$0.310$} & \multirow{4}{*}{$0.512$} & \multirow{2}{*}{$1.459$} & \multirow{2}{*}{$0.789$} & \multirow{2}{*}{$0.853$} \\ \cline{5-6}
&&&& $0.568$ & $0.815$ &   &  &  &    &  &  &   \\ \cline{5-6} \cline{11-13}
&&&& $0.432$ & $0.815$ &   &  &  &    & \multirow{2}{*}{$0.541$} & \multirow{2}{*}{$0.211$} & \multirow{2}{*}{$0.147$}  \\ \cline{5-6}
&&&& $0.568$ & $0.185$ &   &  &  &    &  &  &   \\  \cline{1-1} \cline{4-13}

\multirow{8}{*}{$U_{I, 9}$}& &  & \multirow{4}{*}{NO}
& $0.432$ & $0.815$ & \multirow{4}{*}{$24.186$}  & \multirow{4}{*}{$0.0223$} & \multirow{4}{*}{$0.310$} & \multirow{4}{*}{$0.489$}   & \multirow{2}{*}{$1.541$} & \multirow{2}{*}{$0.209$} & \multirow{2}{*}{$0.146$}  \\ \cline{5-6}
&&&& $0.568$ & $0.185$ &   &  &  &    &  &  &   \\ \cline{5-6} \cline{11-13}
&&&& $0.432$ & $0.185$ &   &  &  &    & \multirow{2}{*}{$0.459$} & \multirow{2}{*}{$0.791$} & \multirow{2}{*}{$0.854$}  \\ \cline{5-6}
&&&& $0.568$ & $0.815$ &   &  &  &    &  &  &   \\  \cline{4-13}

&&&\multirow{4}{*}{IO} & $0.432$ & $0.815$ & \multirow{4}{*}{$27.017$} & \multirow{4}{*}{$0.0226$} & \multirow{4}{*}{$0.310$} & \multirow{4}{*}{$0.488$} & \multirow{2}{*}{$1.541$} & \multirow{2}{*}{$0.211$} & \multirow{2}{*}{$0.147$} \\ \cline{5-6}
&&&& $0.568$ & $0.185$ &   &  &  &    &  &  &   \\ \cline{5-6} \cline{11-13}
&&&& $0.432$ & $0.185$ &   &  &  &    & \multirow{2}{*}{$0.459$} & \multirow{2}{*}{$0.789$} & \multirow{2}{*}{$0.853$}  \\ \cline{5-6}
&&&& $0.568$ & $0.815$ &   &  &  &    &  &  &   \\  \hline \hline
\end{tabular}
\caption{\label{tab:bf_UI_n=4}Results of the $\chi^2$ analysis for the phenomenologically viable mixing patterns $U_{I, 6}$ and $U_{I, 9}$ with $\varphi_1=\pi/4$ and $\varphi_2=\pi/2$. Since the global fit results of the mixing angles slightly differ for normal ordering (NO) and inverted ordering (IO) neutrino mass spectrums, we consider these two cases separately. The $\chi^2$ function reaches a global minimum $\chi^2_{\text{min}}$ at the best fit values $(\theta_{e}, \theta_{\nu})=(\theta^{\text{bf}}_{e}, \theta^{\text{bf}}_{\nu})$. We display the values of the mixing angles and CP violation phases at the best fitting points. }
\end{table}

Regarding the quark mixing, it is described by the CKM matrix which is
parameterized by three mixing angles $\theta^{q}_{ij}$ and one CP-violating phase $\delta^{q}_{CP}$~\cite{Tanabashi:2018oca}.
The full fit values of quark mixing angles and Jarlskog invariant, given by the UTfit collaboration~\cite{utfit:2018}, read as
\begin{eqnarray}
\label{eq:full_fit}
\nonumber && \sin\theta^{q}_{12}=0.22500\pm0.00100, \qquad \sin\theta^{q}_{13}=0.003675\pm0.000095\,, \\
&& \sin\theta^{q}_{23}=0.04200\pm0.00059, \qquad J^{q}_{CP}=(3.120\pm0.090)\times10^{-5}\,.
\end{eqnarray}
Similar to the lepton sector,
the precisely measured CKM mixing matrix can only be explained by the residual symmetry $Z_{2}^{g_{u}}=Z_{2}^{SR^{z_{u}}}$, $X_{u}=\{R^{-z_{u}+x},SR^{x}\}$, $Z_{2}^{g_{d}}=Z_{2}^{SR^{z_{d}}}$ and $X_{d}=\{R^{-z_{d}+y},SR^{y}\}$ where $z_{u}, z_{d}=0,1,\dots,n-1$, $x=y=0$ for odd $n$ and $x,y=0, n/2$ for even $n$. The CKM matrix is predicted to be of the same form as Eq.~\eqref{eq:U_PMNSFin_Z2CPZ2CP} and it reads
\begin{equation}
\label{eq:V_CKM_Z2CPZ2CP}
V_{I}=\left( \begin{array}{ccc}
\cos\varphi_{1} & -c_{d}\sin\varphi_{1} & s_{d}\sin\varphi_{1} \\
c_{u}\sin\varphi_{1} & s_{d}s_{u}e^{i\varphi_{2}}+c_{d}c_{u}\cos\varphi_{1} & c_{d}s_{u}e^{i\varphi_{2}}-s_{d}c_{u}\cos\varphi_{1}\\
-s_{u}\sin\varphi_{1} & s_{d}c_{u}e^{i\varphi_{2}}-c_{d}s_{u}\cos\varphi_{1} & c_{d}c_{u}e^{i\varphi_{2}}+s_{d}s_{u}\cos\varphi_{1} \end{array}\right)\,.
\end{equation}
The values of the parameters $\varphi_1$ and $\varphi_2$ are summarized in table~\ref{Tab:values_of_discrete_parameters_lepton_Z2Z2} where $z_e$ and $z_{\nu}$ should be replaced by $z_{u}$ and $z_d$ respectively. The parameter $\varphi_2$ can be 0, $\pi/2$, $\pi$ and $3\pi/2$, and it should be $\pi/2$ or $3\pi/2$ to explain the observed CP violation in the quark sector.
We have numerically analyzed all the $D_n$ groups with $n\leq40$ and find that the experimentally measured quark mixing matrix can only be described by the following two permutations
\begin{equation}
\label{eq:CKM_possible_caseI}V_{I,1}=V_{I},\qquad V_{I,2}=V_{I}P_{12}\,.
\end{equation}
Accordingly the smallest value of group index $n$ which can accommodate the experimental data is $n=14$. The expressions of the mixing parameters can be read from table~\ref{tab:mixing_parameter_I} by replacing $\theta_{e}$ and $\theta_{\nu}$ with $\theta_{u}$ and $\theta_{d}$ respectively.

For the mixing pattern $V_{I, 1}$, the correlations in Eq.~\eqref{eq:correlation_Ia} are also satisfied for the quark mixing angles $\theta^{q}_{ij}$ and Jarlskog invariant $J^{q}_{CP}$. Moreover we can express the quark CP violation phase $\delta^{q}_{CP}$ in terms of mixing angles,
\begin{equation}\label{eq:pre_DCP_VI_1}
\sin\delta^{q}_{CP}\simeq\frac{\sin2\varphi_{1}}{\sin2\theta^{q}_{12}\cos^2\theta^{q}_{13}\cos\theta^{q}_{23}}\,.
\end{equation}
If we assign the quark doublets to $\mathbf{1}_1\oplus\mathbf{2}_{1}$ of the $D_{14}$ group and choose the residual symmetry $x=7$, $y=0$, $z_d=1$, $z_u=0$, we have $\varphi_{1}=\pi/14$, $\varphi_{2}=\pi/2$ and the observed quark flavor mixing parameters can be accommodated for certain choices of the free parameters $\theta_{u, d}$,
\begin{equation}
\begin{aligned}
&\theta_{u}=0.01237\pi,~~\quad \theta_{d}=0.99473\pi,~~\quad \sin\theta_{12}^{q}=0.22249\,,\\
&\sin\theta_{13}^{q}=0.00369,\quad \sin\theta_{23}^{q}=0.04206,\quad J^{q}_{CP}=3.104\times 10^{-5}\,.
\end{aligned}
\end{equation}
We see that $\sin\theta_{13}^{q}$, $\sin\theta_{23}^{q}$ and $J^{q}_{CP}$ are compatible with the global fitting results of the UTfit collaboration~\cite{utfit:2018}.
The mixing angle $\sin\theta_{12}^{q}$ is about $1\%$ smaller than its measured value and it could be brought into agreement with experimental data in an explicit model since small higher order corrections are generally expected to arise.

For the mixing pattern $V_{I, 2}$, the sum rules in Eqs.~(\ref{eq:correlation_Ib}, \ref{eq:pre_DCP_VI_1}) are fulfilled. The breaking of the flavor group $D_{14}$ and CP symmetry to the residual symmetry with $x=7$, $y=0$, $z_d=6$, $z_u=0$ gives $\varphi_{1}=3\pi/7$, $\varphi_{2}=\pi/2$, we can find values of $\theta_{u, d}$ for which the hierarchical quark mixing angles as well as realistic CP violation phase can be achieved, e.g.
\begin{equation}
\begin{aligned}
&\theta_{u}=0.01326\pi,~~\quad \theta_{d}=0.00117\pi,~~\quad \sin\theta_{12}^{q}=0.22252\,, \\
&\sin\theta_{13}^{q}=0.00357,\quad \sin\theta_{23}^{q}=0.04166,\quad J^{q}_{CP}=3.223\times 10^{-5}\,.
\end{aligned}
\end{equation}
We see that $\sin\theta_{23}^{q}$ is in the experimentally preferred region, the relative deviations of $\sin\theta_{12}^{q}$, $\sin\theta_{13}^{q}$ and $J^{q}_{CP}$ from their best fit values are about $1\%$. This tiny discrepancy should be easily resolved by higher order corrections or renormalization group evolution effects.

Furthermore, the flavor group $D_{14}$ in combination with CP symmetry can reproduce the experimentally favored values of lepton mixing angles if it is broken down to $Z_{2}^{g_{e}}\times X_{e}$ and $Z_{2}^{g_{\nu}}\times X_{\nu}$ in charged lepton sector and neutrino sector respectively. We could choose the residual symmetry specified by $x=7$, $y=0$, $z_{\nu}=4$ and $z_{e}=0$, then the discrete parameters are $\varphi_1=2\pi/7$ and $\varphi_2=\pi/2$. The mixing pattern $U_{I, 9}$ can accommodate the three lepton mixing angles very well, and the best fit values of the mixing parameters are
\begin{equation}
\begin{aligned}
&\theta^{\mathrm{bf}}_{e}=0.439\pi,\quad \theta^{\mathrm{bf}}_{\nu}=0.814\pi,\quad \chi^2_{\text{min}}=1.841,\quad \sin^2\theta_{13}=0.0224,\quad \sin^2\theta_{12}=0.310\,, \\
&\sin^2\theta_{23}=0.602,\quad \delta_{CP}/\pi=1.532,\quad \alpha_{21}/\pi=0.167~(\text{mod}~1),\quad \alpha_{31}/\pi=0.116~(\text{mod}~1)\,.
\end{aligned}
\end{equation}
We would like to remind the readers that the smallest flavor group is $\Delta(294)$ which can accommodate quark and lepton flavor mixing simultaneously if both left-handed quarks and leptons are assigned to an irreducible triplet of $G_f$ and the residual symmetries are $Z_2\times CP$~\cite{Li:2017abz}. Hence the singlet plus doublet assignment seems better than the triplet assignment after including CP symmetry, the order of the flavor symmetry group can be reduced considerably, i.e. 28 versus 294 in this scheme. In particular, the simple dihedral group $D_n$ allows for a unified description of quark and lepton mixing. The dihedral group together with the residual symmetry $Z_2\times CP$ indicated above provides an interesting opportunity for model building.

\section{\label{sec:Z2Z2CP}Mixing patterns from $D_n$ and CP symmetry breaking to $Z_2$ and $Z_2\times CP$ subgroups }
Similar to section~\ref{sec:Z2CPZ2CP}, the left-haded lepton and quark doublets are assigned to the reducible representation $\mathbf{1}_{i}\oplus\mathbf{2}_{j}$, as shown in Eq.~\eqref{eq:assignment}. In this section, we consider the scenario that the residual symmetries of the neutrino and charged lepton mass matrices are $Z_{2}^{g_{\nu}}\times X_{\nu}$ and $Z_{2}^{g_{e}}$ respectively arising from the flavor group $D_n$ and CP. Considering all possible choices for $g_{\nu}$, $X_{\nu}$ and $g_{e}$, we find only the residual symmetry $Z_{2}^{g_{e}}=Z_{2}^{SR^{z_{e}}}$, $Z_{2}^{g_{\nu}}=Z_{2}^{SR^{z_{\nu}}}$ and $X_{\nu}=\{R^{-z_{\nu}+x}, SR^{x}\}$ can lead to mixing pattern in agreement with the present data, where $z_{e}, z_{\nu}=0, 1, \dots, n-1$, $x=0$ if $n$ is an odd integer and $x=0,n/2$ for even $n$. Using the
general formula of Eq.~\eqref{eq:PMNS_Z2CPZ2}, we can get the lepton mixing matrix as follow,
\begin{equation}
\label{eq:U_PMNS_Z2Z2CP}
U_{II}=\left( \begin{array}{ccc}
\cos\varphi_{1} &~ -c_{\nu}\sin\varphi_{1} &~ s_{\nu}\sin\varphi_{1} \\
c_{e}\sin\varphi_{1} &~ s_{\nu}s_{e}e^{i\delta}+c_{\nu}c_{e}\cos\varphi_{1} &~ c_{\nu}s_{e}e^{i\delta}-s_{\nu}c_{e}\cos\varphi_{1}\\
-s_{e}\sin\varphi_{1} &~ s_{\nu}c_{e}e^{i\delta}-c_{\nu}s_{e}\cos\varphi_{1} &~ c_{\nu}c_{e}e^{i\delta}+s_{\nu}s_{e}\cos\varphi_{1} \end{array}\right)\,,
\end{equation}
where $\delta=\varphi_{2}+2\delta_{e}$, $s_{e}=\sin\theta_{e}$, $s_{\nu}=\sin\theta_{\nu}$, $c_{e}=\cos\theta_{e}$, $c_{\nu}= \cos\theta_{\nu}$, the permutation matrices $P_{e, \nu}$ and phase matrices $Q_{e, \nu}$ are omitted. The parameters $\varphi_1$ and $\varphi_2$ are determined by residual symmetry, and their admissible values are summarized in table~\ref{tab:varphi1_varphi2_U2Z2}. We can see that $\varphi_1$ takes the following discrete values
\begin{equation}
\varphi_{1}~(\text{mod}~2\pi)=0,\frac{1}{n}\pi,\frac{2}{n}\pi,\dots,\frac{2n-1}{n}\pi.
\end{equation}
The second discrete parameter $\varphi_2$ appears in $U_{II}$ through the combination $\delta=\varphi_{2}+2\delta_{e}$, the value of $\varphi_2$ is irrelevant since it can be absorbed into the continuous free parameter $\delta_e$. Comparing Eq.~\eqref{eq:U_PMNS_Z2Z2CP} with Eq.~\eqref{eq:U_PMNSFin_Z2CPZ2CP}, we see that $U_{II}$ can be obtained from $U_{I}$ by replacing $\varphi_2$ with $\delta$. Therefore the parameter $\varphi_1$ can be limited in the interval $0\leq\varphi_1\leq\pi/2$, and the variation ranges of the free parameters $\theta_{e}$, $\theta_{\nu}$ and $\delta$ can be taken to be $0\leq\theta_{e}\leq\pi/2$, $0\leq\theta_{\nu}<\pi$ and $0\leq \delta<\pi$ respectively.
\begin{table}[t!]
\centering
\begin{tabular}{|c|c|c|c|c|c|}\hline \hline
$n$ & $i$ & $x$ & $|z_{\nu}-z_{e}|$ &$\varphi_{1}$ & $\varphi_{2}$ \\ \hline
\texttt{odd} & \multirow{3}{*}{$1,2$} & $0$ & &  \multirow{3}{*}{$j(z_{\nu}-z_{e})\pi/n$} & $-j z_{e}\pi/n$ \\
\cline{1-1} \cline{3-3} \cline{6-6}
\multirow{6}{*}{\texttt{even}} &  & $0$ & odd or even & & $-j z_{e}\pi/n$ \\
\cline{3-3} \cline{6-6}
&  & $n/2$  & & & $-j z_{e}\pi/n-j\pi/2$ \\ \cline{2-6}
& \multirow{4}{*}{$3,4$} & \multirow{2}{*}{$0$}  &even &$j(z_{\nu}-z_{e})\pi/n$ & $-j z_{e}\pi/n+z_{\nu}\pi/2$ \\ \cline{4-6}
&  &   & odd &$j(z_{\nu}-z_{e})\pi/n+\pi/2$ & $-j z_{e}\pi/n+(z_{\nu}+1)\pi/2$ \\ \cline{3-6}
&  & \multirow{2}{*}{$n/2$}  & even &$j(z_{\nu}-z_{e})\pi/n$ & $-j z_{e}\pi/n+(z_{\nu}-j-n/2)\pi/2$ \\\cline{4-6}
&  &  &  odd &$j(z_{\nu}-z_{e})\pi/n+\pi/2$ & $-j z_{e}\pi/n+(z_{\nu}-j-n/2+1)\pi/2$ \\ \hline \hline
\end{tabular}
\caption{\label{tab:varphi1_varphi2_U2Z2}The values of the parameters $\varphi_{1}$ and $\varphi_{2}$ for the residual symmetry
$Z_{2}^{g_{e}}=Z_{2}^{SR^{z_{e}}}$, $Z_{2}^{g_{\nu}}=Z_{2}^{SR^{z_{\nu}}}$ and $X_{\nu}=\{R^{-z_{\nu}+x},SR^{x}\}$, where $i$ and $j$ are the indices of the $D_n$ representations $\mathbf{1}_{i}$ and $\mathbf{2}_j$ respectively.}
\end{table}

In this approach, we can not make any prediction for the lepton masses, consequently the lepton mixing matrix is determined up to independent row and column permutations. The residual symmetry fixes one element of the PMNS mixing matrix is $\cos\varphi_{1}$ and it can be any entry. As a result, the 36 possible permutations of rows and columns give rise to nine independent mixing patterns
\begin{equation}
\label{eq:PMNS_caseII}
\begin{aligned}
&U_{II,1}=U_{II}, ~~~ \quad\qquad U_{II,2}=U_{II}P_{12}, ~~~~\quad\qquad U_{II,3}=U_{II}P_{13}\,,\\
&U_{II,4}=P_{12}U_{II},~~\qquad U_{II,5}=P_{12}U_{II}P_{12}, ~~\qquad  U_{II,6}=P_{12}U_{II}P_{13}\,,\\
&U_{II,7}=P_{23}P_{12}U_{II}, ~\quad U_{II,8}=P_{23}P_{12}U_{II}P_{12},~\quad U_{II,9}=P_{23}P_{12}U_{II}P_{13}\,.
\end{aligned}
\end{equation}
For each mixing pattern $U_{II,i}$ ($i=1,2,\dots,9$), the expressions of the mixing parameters can be obtained from those of $U_{I, i}$ in table~\ref{tab:mixing_parameter_I} by replacing $\varphi_{2}$ with $\delta$. From the modulus of the fixed element, we can obtain the following sum rules among the mixing angles and Dirac CP phase,
\begin{eqnarray}
\nonumber&& U_{II, 1}~:~~\cos^{2}\theta_{12}\cos^2\theta_{13}=\cos^2\varphi_{1}\,,\qquad  U_{II, 2}~:~~\sin^{2}\theta_{12}\cos^2\theta_{13}=\cos^2\varphi_{1}\,, \\
\nonumber && U_{II, 6}~:~~\sin^{2}\theta_{23}\cos^2\theta_{13}=\cos^2\varphi_{1}\,,\qquad  U_{II, 9}~:~~\cos^{2}\theta_{23}\cos^2\theta_{13}=\cos^2\varphi_{1}\,, \\
\nonumber&& U_{II, 4}~:~~\cos\delta_{CP}=\frac{2(\cos^2\varphi_{1}-\sin^2\theta_{12}\cos^2\theta_{23}-\sin^2\theta_{13}\cos^2\theta_{12}\sin^2\theta_{23})}
{\sin2\theta_{12}\sin\theta_{13}\sin2\theta_{23}}\,, \\
\nonumber&& U_{II, 5}~:~~\cos\delta_{CP}=-\frac{2(\cos^2\varphi_{1}-\cos^2\theta_{12}\cos^2\theta_{23}-\sin^2\theta_{13}\sin^2\theta_{12}\sin^2\theta_{23})}
{\sin2\theta_{12}\sin\theta_{13}\sin2\theta_{23}}\,, \\
\nonumber&& U_{II, 7}~:~~\cos\delta_{CP}=-\frac{2(\cos^2\varphi_{1}-\sin^2\theta_{12}\sin^2\theta_{23}-\sin^2\theta_{13}\cos^2\theta_{12}\cos^2\theta_{23})}
{\sin2\theta_{12}\sin\theta_{13}\sin2\theta_{23}}\,, \\
\label{eq:correlation_I}&& U_{II, 8}~:~~\cos\delta_{CP}=\frac{2(\cos^2\varphi_{1}-\cos^2\theta_{12}\sin^2\theta_{23}-\sin^2\theta_{13}\sin^2\theta_{12}\cos^2\theta_{23})}
{\sin2\theta_{12}\sin\theta_{13}\sin2\theta_{23}}\,.
\end{eqnarray}
In order to show concrete examples and find new interesting mixing, we have numerically scanned over the free parameters $\theta_{e}$, $\theta_{\nu}$ and $\delta$ and all possible values of the discrete parameter $\varphi_1$ for each integer group index $n$.
We find the smallest dihedral group which can accommodate
the data is $D_3\cong S_3$. Note that $D_3$ group with $n=3$ is the symmetry group of an equilateral triangle, and then $\varphi_1$ can be either $0$ or $\pi/3$ in the fundamental region of  $\varphi_{1}\in[0,\pi/2]$. Only the value $\varphi_{1}=\pi/3$ can generate a viable mixing pattern, and it can be achieved from the residual symmetry $g_e=S$, $g_{\nu}=SR$, $X_{\nu}=\left\{R^2, S\right\}$ under the lepton doublets assignment $\mathbf{1}_1\oplus\mathbf{2}_1$. Accordingly the fixed element is $\cos\varphi_{1}=1/2$ and it can be the $(21)$, $(22)$, $(31)$ or $(32)$ entry of the lepton mixing matrix. Hence only the mixing patterns $U_{II,4}$, $U_{II,5}$, $U_{II,7}$ and $U_{II,8}$
can be compatible with experimental data.
Requiring all the three mixing angles $\theta_{12}$, $\theta_{13}$ and $\theta_{23}$ in the $3\sigma$ intervals of global fit~\cite{Esteban:2018azc}, we can obtain the allowed regions of the mixing angles and CP violation phases and the numerical results are summarized in table~\ref{Tab:z2z2cp_lepton_v2}.
\begin{table}[t!]
\centering
\footnotesize
\begin{tabular}{|c |c c c c c c|}
\hline \hline
   & $\theta_{13}/^{\circ}$ & $\theta_{12}/^{\circ}$ & $\theta_{23}/^{\circ}$ & $\delta_{CP}/\pi$ & $\alpha_{21}/\pi~(\text{mod}~1)$ & $\alpha_{31}/\pi~(\text{mod}~1)$ \\
\hline
  \multirow{2}{*}{$U_{II,4}\big|_{\varphi_1=\frac{\pi}{3}}$} & \multirow{2}{*}{$8.220-8.981$} & \multirow{2}{*}{$33.258-36.271$} & \multirow{2}{*}{$40.9-46.246$} & $0-0.299 $ & $0-0.139$ & $0-0.085$ \\
     &&&&$ \oplus 1.701-2$&$ \oplus 0.861-1$& $ \oplus 0.915-1$\\
\hline
  \multirow{2}{*}{$U_{II,5}\big|_{\varphi_1=\frac{\pi}{3}}$} & \multirow{2}{*}{$8.220-8.981$} & \multirow{2}{*}{$31.628-36.271$} & \multirow{2}{*}{$45.453-52.180$} & $0-0.505 $ & $0-0.263 $ & $0-0.172 $ \\
       &&&&$ \oplus 1.495-2$&$ \oplus 0.737-1$& $\oplus 0.828-1$\\
\hline
  \multirow{2}{*}{$U_{II,7}\big|_{\varphi_1=\frac{\pi}{3}}$} & \multirow{2}{*}{$8.220-8.981$} & \multirow{2}{*}{$32.032-36.271$} & \multirow{2}{*}{$43.766-52.180$} & \multirow{2}{*}{$0.617-1.383$} & $0-0.147$ & $0-0.092$ \\
     &&&&&$ \oplus 0.853-1$& $ \oplus 0.908-1$\\
\hline
  \multirow{2}{*}{$U_{II,8}\big|_{\varphi_1=\frac{\pi}{3}}$} & \multirow{2}{*}{$8.220-8.981$} & \multirow{2}{*}{$31.628-36.271$} & \multirow{2}{*}{$40.9-44.542$} & \multirow{2}{*}{$0.650-1.350 $} & $0-0.202$ & $0-0.124$ \\
     &&&&&$ \oplus 0.798-1$& $ \oplus 0.877-1$\\
  \hline \hline
\end{tabular}
\caption{\label{Tab:z2z2cp_lepton_v2}The allowed ranges of the mixing parameters for the mixing patterns $U_{II,4}$, $U_{II,5}$, $U_{II,7}$ and $U_{II,8}$. Here we choose $\varphi_1=\pi/3$ which is the unique viable value of $\varphi_1$ in the $D_3$ flavor group.}
\end{table}

Subsequently we extend the above scheme to the quark sector, the flavor symmetry $D_n$ and CP symmetry are broken down to $Z_{2}^{g_{u}}$ and $Z_{2}^{g_{d}}\times X_{d}$ in the up quark and down quark sectors respectively. The CKM mixing matrix can be correctly reproduced if the residual symmetry is $g_u=SR^{z_{u}}$, $g_d=SR^{z_{d}}$ and $X_{d}=\{R^{-z_{d}+x},SR^{x}\}$ with $z_{u}, z_{d}=0,1,\dots, n-1$, $x=0$ for odd $n$ and $x=0, n/2$ for even $n$. The CKM matrix is determined to be
\begin{equation}
\label{eq:U_CKM_Z2Z2CP_form}
V_{II}=\left(\begin{array}{ccc}
\cos\varphi_{1} &~ -c_{d}\sin \varphi_{1} ~& s_{d}\sin \varphi_{1}\\
c_{u}\sin \varphi_{1} &~ s_{u} s_{d}e^{i \delta} +c_{u} c_{d}\cos \varphi_{1} ~& s_{u} c_{d}e^{i \delta} -s_{d} c_{u}\cos \varphi_{1} \\
-s_{u}\sin \varphi_{1} &~ s_{d} c_{u}e^{i \delta} -s_{u} c_{d}\cos \varphi_{1} ~& c_{u} c_{d}e^{i \delta}+s_{u} s_{d}\cos \varphi_{1}
\end{array}
\right)\,,
\end{equation}
up to permutations of rows and columns, where $c_{u}=\cos\theta_{u}$, $c_{d}=\cos\theta_{d}$, $s_{u}=\sin\theta_{u}$, $s_{d}=\sin\theta_{d}$ and $\delta=\varphi_{2}+2\delta_{u}$. The values of the discrete parameters $\varphi_{1}$ and $\varphi_{2}$ can be read from table~\ref{tab:varphi1_varphi2_U2Z2} by substituting $z_{e}$ and $z_{\nu}$ with $z_{u}$ and $z_{d}$ respectively. Furthermore, it is straightforward to check that the same mixing pattern would be obtained if the residual symmetry is instead $Z^{g_{u}}_2=Z_{2}^{SR^{z_{u}}}$, $X_{u}=\{R^{-z_{u}+x}, SR^{x}\}$ and $Z^{g_{d}}_2=Z_{2}^{SR^{z_{d}}}$.

Similar to the lepton mixing matrices in Eq.~\eqref{eq:PMNS_caseII}, the row and columns permutations of $V_{II}$ can give rise to nine mixing patterns $V_{II, i}$ ($i=1,\ldots,9$). The mixing matrix $V_{II, i}$ can be obtained from $U_{II, i}$ by replacing $\theta_{e}$, $\theta_{\nu}$ and $\delta_{e}$ with $\theta_{u}$, $\theta_{d}$ and $\delta_u$ respectively. We have considered all possible values of the discrete parameters $\varphi_{1}$ for each group index $n$ with $n\leq 40$. We scan over the free parameters $\theta_{u}$, $\delta_{u}$ and $\theta_{d}$ in the range from $0$ and $\pi$ to determine whether a good fit to the experimental data can be achieved.
For the $D_{n}$ groups with $n\leq40$, we find six permutations $V_{II,1}$, $V_{II,2}$, $V_{II,4}$, $V_{II,5}$, $V_{II,6}$, and $V_{II,8}$ can describe the measured values of the quark mixing parameters shown in Eq.~\eqref{eq:full_fit}. The values of $n$, $\varphi_{1}$ and the resulting predictions for $\sin\theta_{ij}^{q}$ and $J^{q}_{CP}$ at certain benchmark values of $\theta_{u}$, $\delta_{u}$, $\theta_{d}$ are summarized in table~\ref{Tab:numerical_result_quark_z2z2cp}. We see that the smallest group index $n$ which can accommodate the experimental data is $n=7$ and accordingly the mixing patterns are $V_{II,2}$ and $V_{II,4}$.
\begin{table}[t!]
\centering
\footnotesize
\begin{tabular}{|c|c|c|c|c|c|c|c|c|c|}
\hline\hline
& $n$ & $\varphi_{1}$ & $\theta_{u}/\pi$ & $\theta_{d}/\pi$  & $\delta/\pi$ & $\sin\theta_{13}^{q}$ & $\sin\theta_{12}^{q}$ & $\sin\theta_{23}^{q}$ & $J^{q}_{CP}/10^{-5}$ \\
\hline
$V_{II,1}$ & $14,28$ & $\pi/14$ & $0.98405$ & $0.00526$  & $0.71399$ & $0.00367$ & $0.22249$ & $0.04200$ & $3.120$ \\ \hline
$V_{II,2}$ & $7,14,21,28,35$ & $3\pi/7$ & $0.01347$  & $0.00120$  & $0.37658$ & $0.00368$ & $0.22252$ & $0.04200$ & $3.120$ \\ \hline
$V_{II,4}$ & $7,14,21,28,35$ & $3\pi/7$ & $0.00365$ & $0.01372$  & $0.09649$ & $0.00368$ & $0.22277$ & $0.04201$ & $3.118$ \\  \hline
$V_{II,5}$ & $27$ & $2\pi/27$ & $0.94259$  & $0.05847$ & $0.99395$ & $0.00368$ & $0.22688$ & $0.04212$ & $3.119$  \\  \hline
\multirow{2}{*}{$V_{II,6}$} & $37$ & $18\pi/37$ & $0.50117$ & $0.07226$  & $0.36627$ & $0.00367$ & $0.22500$ & $0.04244$ & $3.120$  \\ \cline{2-10}
& $39$ & $19\pi/39$ & $0.50117$ & $0.07225$ & $0.41240$  & $0.00368$ & $0.22500$ & $0.04027$ & $3.120$ \\ \hline
\multirow{2}{*}{$V_{II,8}$}
& $37$ & $18\pi/37$ & $0.42770$  & $0.49723$ & $0.12558$ & $0.00369$ & $0.22499$ & $0.04316$ & $3.108$ \\\cline{2-10}
& $39$ & $19\pi/39$ & $0.42770$  & $0.49643$ & $0.10190$ & $0.00368$ & $0.22502$ & $0.04164$ & $3.114$ \\ \hline\hline
\end{tabular}
\caption{\label{Tab:numerical_result_quark_z2z2cp} Numerical results of the quark mixing parameters for the permutations of the mixing matrix $V_{II}$ in Eq.~\eqref{eq:U_CKM_Z2Z2CP_form}, where the residual symmetry is $Z_{2}^{g_{u}}=Z^{SR^{z_{u}}}_2$, $Z_{2}^{g_{d}}=Z^{SR^{z_{d}}}_2$, $X_{d}=\{R^{-z_{d}+x},SR^{x}\}$. We have analyzed all the $D_{n}$ groups with $n\leq40$. Here we show the values of $\sin\theta_{ij}^{q}$ and $J_{CP}^{q}$ which are compatible with the experimental data for certain choices of $\theta_{u}$, $\delta_{u}$, $\theta_{d}$ and $\varphi_{1}$.}
\end{table}

\begin{figure}[t!]
\centering
\includegraphics[width=0.98\textwidth]{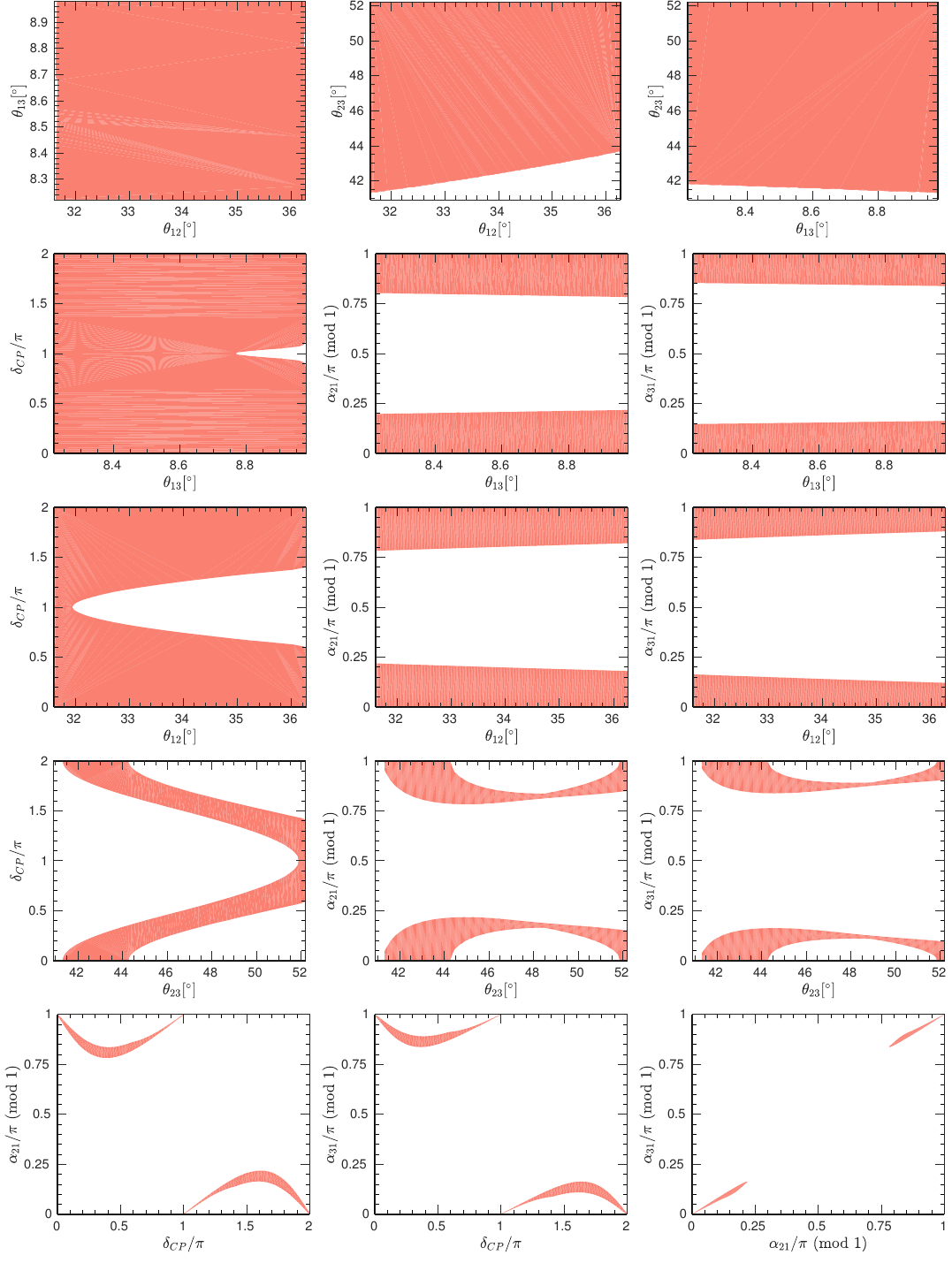}
\caption{\label{fig:U_II_8_n_7_correlations}Correlations between different mixing parameters for the mixing pattern $U_{II,8}$ with $\varphi_1=2\pi/7$, and the three lepton mixing angles are required to lie in their $3\sigma$ ranges~\cite{Esteban:2018azc}.}
\end{figure}

Furthermore, we notice that the $D_7$ group and CP symmetry can also generate phenomenologically viable lepton mixing patterns if the residual symmetries of the charged lepton and neutrino mass matrices are $Z_{2}^{g_{e}}=Z_{2}^{SR^{z_{e}}}$ and $Z_{2}^{g_{\nu}}=Z_{2}^{SR^{z_{\nu}}}$, $X_{\nu}=\{R^{-z_{\nu}+x}, SR^{x}\}$ respectively. We find that only the mixing patterns $U_{II,4}$, $U_{II,5}$, $U_{II,8}$ and $U_{II,9}$ can agree well with the experimental data on lepton mixing angles, and the discrete parameter $\varphi_1$ can be $2\pi/7$ or $3\pi/7$. The continuous parameters $\theta_{e}$, $\delta$ and $\theta_{\nu}$ are freely varied between 0 and $\pi$, and the current $3\sigma$ bounds of $\sin^2\theta_{ij}$~\cite{Esteban:2018azc} are imposed. The allowed regions of the lepton mixing angles and CP phases are reported in table~\ref{Tab:z2z2cp_lepton_n=7}. As an example, we display the correlations among the different mixing parameters for $U_{II,8}\big|_{\varphi_1=\frac{2\pi}{7}}$ in figure~\ref{fig:U_II_8_n_7_correlations}. Comparing with the scenario of $Z_2\times CP$ residual symmetry in both neutrino and charged lepton sectors, we see that the admissible region of the Dirac phase $\delta_{CP}$ is generally more larger. It is remarkable that the $D_7$ flavor symmetry with group order 14 already can give experimentally favored values of PMNS and CKM matrix. For the irreducible triplet assignment of quark and lepton doublets, we would like to mention that $\Delta(294)$ is the minimal flavor group to generate realistic quark and lepton flavor mixing patterns in the present scheme~\cite{Lu:2018oxc}.
\begin{table}[t!]
\centering
\footnotesize
\begin{tabular}{|c|cccccc|}
\hline \hline
case & $\theta_{13}/^{\circ}$ & $\theta_{12}/^{\circ}$ & $\theta_{23}/^{\circ}$ & $\delta_{CP}/\pi$ & $\alpha_{21}/\pi~(\text{mod}~1)$ & $\alpha_{31}/\pi~(\text{mod}~1)$ \\\hline
\multirow{2}{*}{$U_{II,4}\big|_{\varphi_1=\frac{3\pi}{7}}$} & \multirow{2}{*}{$8.472-8.981$} & \multirow{2}{*}{$31.628-32.195$} & \multirow{2}{*}{$51.502-52.180$} & \multirow{2}{*}{$0.912-1.088 $} & $0-0.101$ & $0-0.082$ \\
&&&&&$ \oplus 0.899-1$& $ \oplus 0.918-1$\\\hline
\multirow{2}{*}{$U_{II,5}\big|_{\varphi_1=\frac{2\pi}{7}}$} & \multirow{2}{*}{$8.220-8.981$} & \multirow{2}{*}{$31.628-36.271$} & \multirow{2}{*}{$40.9-48.673$} & \multirow{2}{*}{$0.353-1.646$} & $0-0.218 $ & $0-0.163 $ \\
&&&&&$ \oplus 0.782-1$& $\oplus 0.837-1$\\\hline
\multirow{2}{*}{$U_{II,8}\big|_{\varphi_1=\frac{2\pi}{7}}$} & \multirow{2}{*}{$8.220-8.981$} & \multirow{2}{*}{$31.628-36.271$} & \multirow{2}{*}{$41.307-52.180$} & \multirow{2}{*}{$0-2$} & $0-0.218$ & $0-0.163 $ \\
&&&&&$ \oplus 0.782-1$& $\oplus 0.837-1$\\\hline
\multirow{2}{*}{$U_{II,9}\big|_{\varphi_1=\frac{2\pi}{7}}$} & \multirow{2}{*}{$8.223-8.981$} & \multirow{2}{*}{$31.628-36.271$} & \multirow{2}{*}{$50.860-50.952$} & \multirow{2}{*}{$0-2 $} & $0-0.181 $ & $0-0.132 $ \\
&&&&&$ \oplus 0.819-1$& $\oplus 0.868-1$\\\hline \hline
\end{tabular}
\caption{\label{Tab:z2z2cp_lepton_n=7}
The allowed ranges of the mixing parameters for the viable mixing patterns $U_{II,4}$, $U_{II,5}$, $U_{II,8}$ and $U_{II,9}$, and the flavor symmetry group is $D_7$ such that the viable values of $\varphi_1$ are $2\pi/7$ and $3\pi/7$.}
\end{table}

\section{\label{sec:summary}Summary and conclusions }

A compelling theory of flavor mixing is still missing. The discrete flavor symmetry and CP symmetry through the mismatch of residual symmetries is a powerful approach to explain the observed flavor mixing structure of quarks and leptons. In previous work, we find that realistic CKM and PMNS matrices can be achieved if the residual symmetry in the neutrino and down quark sectors is $Z_2\times CP$, and a subgroup $Z_2\times CP$ or $Z_2$ is preserved by the charged lepton and up quark mass matrices~\cite{Li:2017abz,Lu:2018oxc}. If the three generations of left-handed quark and lepton doublets transform as an irreducible three-dimensional representation of the flavor symmetry group, the minimal group turns out to be $\Delta(294)$~\cite{Li:2017abz,Lu:2018oxc}. The motivation of the present work is to find a smaller flavor group which can give a unified description of quark and lepton flavor mixing.

In this paper, we perform a detailed analysis of the dihedral group $D_n$ as flavor symmetry in combination with CP symmetry. We have identified the most general form of the CP transformations compatible with $D_n$. Since the group $D_n$ only has one-dimensional and two-dimensional irreducible representations, the left-handed quark and lepton fields are assigned to the direct sum of a singlet and a doublet of $D_n$. If the symmetries $D_n$ and CP are broken in such a way that neutrino and charged lepton sectors remain invariant under two different $Z_2\times CP$ subgroups, all the lepton mixing angles and CP phases would depend on only two real free parameters $\theta_{e}$ and $\theta_{\nu}$. The measured values of the lepton mixing angles can be explained by small group $D_4$ which is the symmetry group of a square, see table~\ref{tab:bf_UI_n=4} for numerical results. In the same way as presented for leptons, viable quark mixing can be derived under the assumption that the residual symmetries of the up and down quark sectors are $Z_2\times CP$ as well. Moreover, we find that the flavor group $D_{14}$ can give the experimentally favored CKM and PMNS mixing matrices.

Furthermore, we consider a second scenario in which the residual symmetries of the charged lepton and up quark sectors are $Z_2$ instead of $Z_2\times CP$ while the neutrino and down quark mass matrices remain invariant under a $Z_2\times CP$ subgroup. The resulting lepton and quark mixing matrices would depend on three free parameters $\theta_{e}$, $\theta_{\nu}$, $\delta_{e}$ and $\theta_{u}$, $\theta_{d}$, $\delta_{u}$ respectively. The observed patterns of quark and lepton flavor mixing can be accommodated by the $D_7$ group.

In the approach with only flavor symmetry (without CP), in order to achieve at least two non-vanishing mixing angles, the left-handed leptons are usually assumed to transform as an irreducible three-dimensional representation under the flavor symmetry group. An important lesson what we learn is that the singlet plus doublet assignment also allows one to understand the experimental data on lepton mixing angles after the CP symmetry is considered. We conclude that dihedral group and CP symmetry provide new opportunity for building models of quark and lepton mixing. It is interesting to implement the presented symmetry breaking patterns here in a concrete model, and as usual the assumed residual symmetries could be dynamically realized through non-vanishing vacuum expectation values of some
flavons.

\section*{Acknowledgements}

This work is supported by the National Natural Science Foundation of China under Grant Nos.11522546 and 11835013.

\vskip 2cm

\bibliographystyle{utphys}
\bibliography{dihedral_CP}

\end{document}